\documentclass[twoside,compsoc]{IEEEtran}
\usepackage{makecell}
\usepackage{hyperref}
\usepackage{array}
\usepackage{graphicx,amssymb,amsmath}
\usepackage{multicol}
\usepackage[noadjust]{cite}
\usepackage{setspace}
\usepackage{float}
\usepackage {url}
\usepackage{stfloats}
\usepackage{amsthm,pifont}
\usepackage{flushend}
\usepackage[table]{xcolor}
\usepackage{multirow}
\usepackage{caption}
\usepackage{rotating}
\usepackage{cases,subeqnarray}
\usepackage{bm,multirow,bigstrut}
\usepackage{textcomp}
\usepackage{latexsym,bm}
\usepackage{booktabs}
\usepackage{xcolor}
\usepackage{mathtools}
\usepackage{dsfont}
\usepackage{extarrows}
\usepackage{epsfig}
\usepackage{epsfig}
\usepackage{epstopdf}
\usepackage{colortbl}
\usepackage{algpseudocode}
\usepackage{algorithmicx,algorithm}
\usepackage[caption=false,font=footnotesize]{subfig}
\theoremstyle{plain}

\theoremstyle{plain}

\usepackage{amsmath}

\definecolor{Gray}{gray}{0.85}

\IEEEoverridecommandlockouts
\begin{document}

\title{MORES: Mobile Reasoning-as-a-Service via Distributed LLM Inference-Time Scaling}

\author{Guanchen Liu, Hongyang Du$^*$, Kaibin Huang
\thanks{G. Liu, H. Du, K. Huang are with the Department of Electrical and Computer Engineering, University of Hong Kong, Hong Kong SAR, China. (email: liugc@connect.hku.hk, duhy@hku.hk, huangkb@hku.hk)}
}
\maketitle
\vspace{-1cm}

\begin{abstract}
Inference-time scaling has emerged as an effective approach for enhancing the capabilities of Large Language Models (LLMs), addressing the growing demand for stronger reasoning without increasing model size.
This novel form of LLM scaling comprises two representative approaches: {\textit{explicit reasoning}}, which generates intermediate chain-of-thought tokens during an explicit thinking phase, and {\textit{implicit reasoning}}, which iteratively updates hidden states in the latent space without producing explicit outputs.
Despite their effectiveness, both paradigms incur substantial computational and memory overhead, raising challenges for deployment on resource-constrained edge devices.
To address these issues, we propose a Mobile Reasoning-as-a-Service (MORES) framework that treats reasoning as a computational service accessible to edge devices over wireless networks.
Focusing on implicit reasoning, we leverage its recursive structure to partition hidden-state updates between edge devices and servers, enabling cooperative inference that allows devices to access additional cloud computation on demand.
To optimize long-term performance, we formulate a joint computation and communication scheduling problem and solve it using a semantic Mixture-of-Experts (MoE)-based Deep Reinforcement Learning (DRL) algorithm to address heterogeneity in wireless conditions and task demands.
The agent adaptively allocates resources by adjusting the number of recurrent steps and the transmission pruning rate, while a semantic router enables high-speed gating for real-time expert selection.
Experimental results show that the proposed method achieves an approximately 18\% improvement in system throughput over the baseline Soft Actor-Critic (SAC) algorithm. Our code is available at https://github.com/NICE-HKU/MORES.
\end{abstract}

\IEEEpeerreviewmaketitle
\section{Introduction}
Large Language Models (LLMs) have scaled rapidly, with parameter counts increasing from 175 billion in OpenAI's GPT-3~\cite{brown2020gpt3} to trillions in models such as GPT-5~\cite{singh2025openaigpt5}, DeepSeek-V4~\cite{deepseekai2026deepseekv4}, and Google's Gemini 3~\cite{google2025gemini3blog2}.
This parameter-driven growth followed predictable scaling laws~\cite{zhou2024large,hoffmann2022empirical}, where larger models outperformed smaller ones across diverse tasks.
However, further expansion faces diminishing returns. Model training becomes prohibitively expensive, computational costs grow quadratically, and performance gains are increasingly constrained by data quality and availability.
To overcome these limitations, researchers have introduced a new paradigm known as inference-time scaling~\cite{google-inference-scaling}. Rather than expanding model parameters, this approach increases computation during inference to enhance reasoning capabilities.
DeepSeek-R1~\cite{guo2025deepseek} represents a breakthrough in this direction by applying Deep Reinforcement Learning (DRL) to train LLMs that generate extensive Chain-of-Thought (CoT) tokens, significantly improving performance on complex tasks like mathematical problem-solving and logical reasoning.
Subsequent efficient reasoning models like QwQ-32B~\cite{qwq32b} further validate that extended reasoning processes at inference time can produce results comparable to much larger models.
This paradigm shift enables smaller LLMs to match or exceed the performance of much larger LLMs by trading parameter count for inference-time computation, offering a more efficient path to advanced models' reasoning capabilities.

\begin{figure}[t]
\centering
\includegraphics[width=1.0\columnwidth]{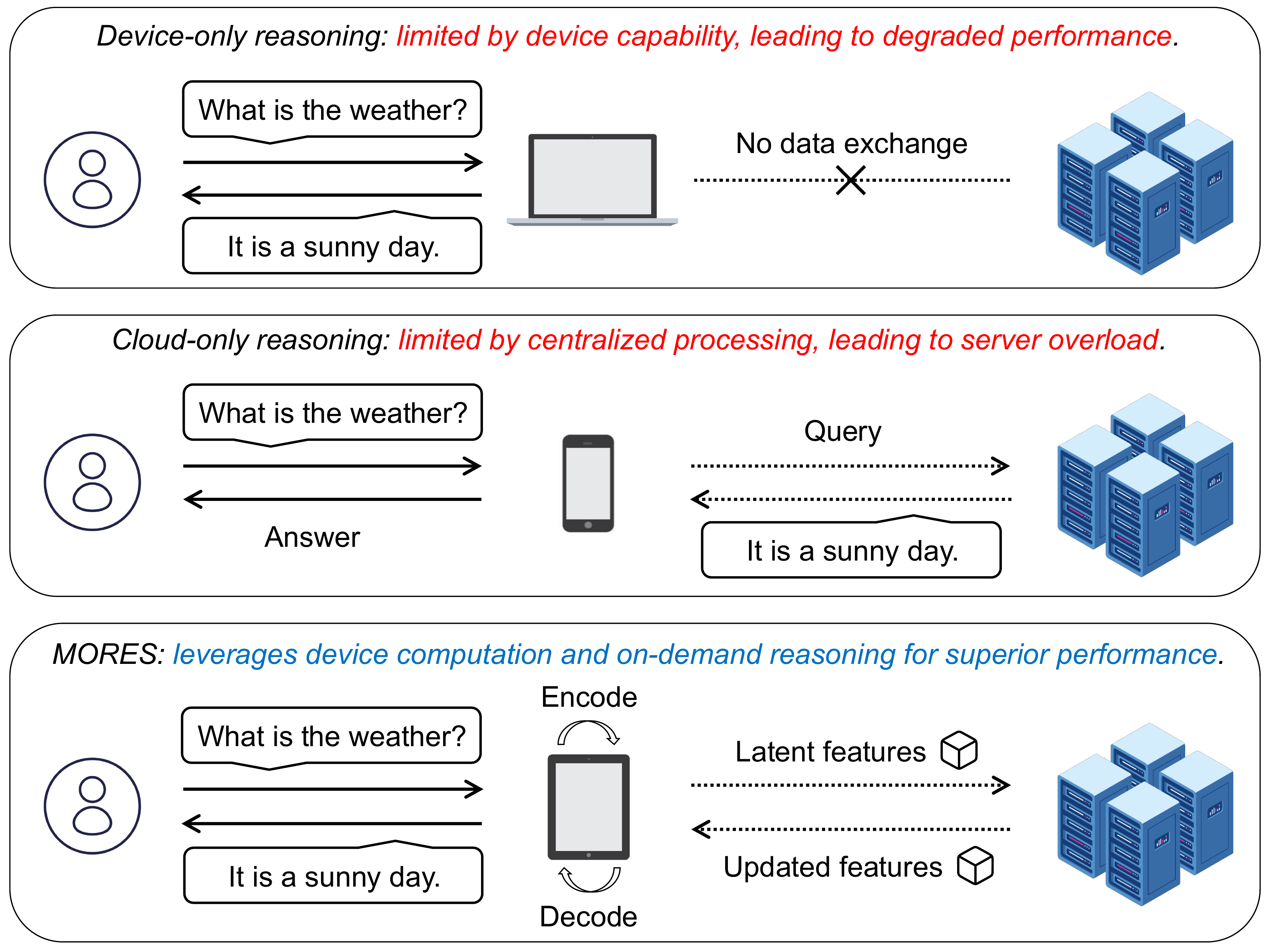}
\caption{Comparison of reasoning modes between device-only, cloud-only, and the proposed device–cloud collaborative MORES architectures.}
\label{fig: intro_model}
\end{figure}

Two distinct approaches to inference-time scaling have emerged recently.
The first approach employs {\textit{explicit reasoning}}, exemplified by DeepSeek-R1~\cite{guo2025deepseek} and OpenAI o1~\cite{jaech2024openai}, where LLMs generate extensive CoT tokens during intermediate reasoning steps before giving the final answer.
The second approach leverages {\textit{implicit reasoning}} in latent space~\cite{geiping2025scaling}, which recursively applies recurrent Transformer blocks to hidden states without generating additional tokens, thereby enabling compact reasoning with lower memory and computation overhead.
While both approaches significantly improve model reasoning capabilities, they pose serious challenges for deployment on resource-constrained devices.
In particular, explicit reasoning models involve processing and storing long token sequences, leading to high memory usage and computational overhead.
In contrast, implicit reasoning is more efficient in token usage but still requires intensive computation due to deep recursive operations over high-dimensional hidden states~\cite{latent2025survey}.
Consequently, these additional computational demands fundamentally complicate system deployment in practice.
As illustrated in Fig.~\ref{fig: intro_model}, performing all reasoning on centralized cloud servers can easily result in server overload under large-scale workloads, while executing the full reasoning process on resource-constrained devices is often infeasible due to limited computation and energy budgets.
This imbalance between computational demand and available system resources severely limits the scalability of existing deployment approaches.

To address these deployment challenges, prior research has investigated model-centric solutions, including fine-tuning LLMs on task-specific reasoning datasets and distilling reasoning capabilities from larger models into smaller ones~\cite{distill-step-by-step}.
However, these approaches mainly aim to reduce model size and computational demand under a centralized inference assumption.
In particular, existing approaches do not explicitly consider distributed and resource-heterogeneous edge networks, where computation can be jointly provided by devices and edge servers.
Beyond the deployment challenges regarding server capacity and device capabilities, another limitation arises from the nature of explicit CoT reasoning.
Under this paradigm, the reasoning process is explicitly represented as intermediate CoT tokens.
This design introduces additional intermediate reasoning tokens beyond the final answer, resulting in extra communication overhead and latency uncertainty over wireless links.
Moreover, recent studies suggest that such an explicit CoT may not faithfully reflect the internal reasoning process and can be unstable across generations.
To overcome these limitations in wireless and resource-constrained environments, we propose a device-edge cooperative reasoning paradigm based on a latent reasoning model to jointly leverage the computing capabilities of edge devices and edge servers.
In this paradigm, edge devices perform initial reasoning locally, while edge servers provide on-demand reasoning computation to extend inference-time reasoning beyond local device capabilities. This cooperative design also enables implicit latent reasoning, thereby reducing communication overhead compared with explicit CoT-based reasoning.
Such a cooperative paradigm imposes two essential system requirements from the computation and communication perspectives:
\begin{itemize}
\item \textit{R1. Computation Requirement.} 
The system should efficiently utilize available computational resources to support inference-time reasoning across heterogeneous computing nodes.
\item \textit{R2. Communication Requirement.} 
The system should provide efficient and reliable information exchange under diverse wireless network conditions to enable coordinated system operation.
\end{itemize}

To meet these requirements, an effective framework should be designed with the following principles:
\begin{itemize}
\item \textit{P1. Computation Principle.}
The framework should adopt a distributed reasoning paradigm across heterogeneous mobile networks, allowing devices to request adjustable reasoning depth from cloud servers on demand based on task complexity and available system resources.
\item \textit{P2. Communication Principle.} 
The framework should enable efficient system coordination by leveraging information compression and selective transmission, thereby reducing communication latency and energy consumption under wireless constraints.
\end{itemize}

Guided by the above requirements and design principles, we propose Mobile Reasoning-as-a-Service (MORES). As illustrated in Fig.~\ref{fig: intro_model}, MORES treats reasoning as a distributed computational service across wireless network nodes.
Specifically, MORES focuses on latent reasoning, since its intermediate latent representation and recursive structure are well suited for partitioning across devices.
In MORES, resource-constrained edge devices perform lightweight local inference and offload intermediate latent representations and states to edge servers, which further update these latent states through deeper reasoning and return the refined latent states to the devices for output generation.
This design preserves reasoning quality while utilizing device-side computation.

In LLM reasoning, the appropriate reasoning depth inherently depends on the query.
For example, simple question-answering tasks typically require only a few reasoning steps, whereas more complex tasks like mathematics or code generation benefit from deeper reasoning.
In MORES, such variability calls for on-demand coordination between edge devices and servers to determine how many reasoning steps should be requested from the server side.
During this process, intermediate latent representations and states are exchanged between edge devices and servers, resulting in additional communication overhead.
To address this, MORES introduces an adaptive pruning mechanism that compresses latent features before transmission, thereby reducing communication latency and energy consumption while preserving the semantic information required for subsequent reasoning.

To achieve both \textit{P1} and \textit{P2}, we formulate a joint decision-making problem and solve it using a semantic Mixture-of-Experts (MoE)-based DRL algorithm.
The main contributions of this paper are as follows:
\begin{itemize}
\item We introduce MORES, a distributed inference-time scaling framework in wireless networks, enabling devices to access on-demand reasoning support with enhanced quality.
\item We leverage the inherent recursive structure of latent LLM reasoning to facilitate natural computation partitioning between edge devices and edge servers. Additionally, the recurrent steps and the pruning rate are jointly optimized to achieve efficient reasoning.
\item We develop a semantic MoE-based DRL algorithm that effectively addresses system heterogeneity. The Soft Actor-Critic (SAC) agent makes adaptive decisions based on varying tasks and wireless conditions to maximize long-term performance.
\end{itemize}

The remainder of this paper is organized as follows: Section~\ref{sec: related_work} discusses related work in LLM inference-time scaling, wireless network-aided LLMs, and MoE-aided DRL.
Section~\ref{sec: system_model} introduces the proposed MORES system model, which integrates the distributed reasoning, computation, and communication models, followed by the formulation of the throughput maximization problem.
Section~\ref{sec: semantic_moe_drl} presents the proposed semantic MoE-based DRL algorithm, including the DRL solution, the semantic router, and the MoE architecture design.
Section~\ref{sec: experiments} presents comprehensive numerical results and performance analysis.
Finally, Section~\ref{sec: conclusion} concludes the paper with a summary of our key findings.
A list of mathematical symbols frequently used in this paper is shown in Table~\ref{table_one}.

\renewcommand{\arraystretch}{1.1}
\begin{table}[t]
\caption{Mathematical Notations}
\label{table_one}
\centering
\begin{tabular}{p{2cm} p{6cm}}
\hline
\textbf{Notation} & \textbf{Description} \\ \hline
$\mathbf{x}$ & Prompt input token sequence \\
$\mathbf{y}$ & Prompt output token sequence \\
$\mathbf{e}$ & Latent representation \\
$\mathbf{z}$ & Latent state \\
$\mathbf{P}$ & Decoding probability scores \\
\hline
$\mathcal{P}$ & Prelude encoder \\
$\mathcal{C}$ & Coda decoder \\
$\mathcal{R}$ & Recurrent reasoning unit \\
\hline
$n$ & The number of input tokens \\
$m$ & The number of output tokens\\
$\alpha$ & The number of tokens processed\\
$r$ & Recurrent steps \\
$l$ & The index of generation round \\
$k$ & The index of request \\
$t$ & Time slot \\
$d$ & Task type\\
$h$ & Channel gain\\
$q$ & Quantization bit width\\
$v$ & Hidden size of latent space \\
$\rho$ & Pruning rate \\
$\sigma_z$ & Latent state noise scale \\
$\sigma_h$ & Channel noise scale \\
$\eta$ & Path loss factor \\
$\delta$ & The correctness of reasoning result \\
$U$ & Average system throughput \\
$K$ & The number of requests \\
$G_{\max}$ & Maximum recurrence budget \\
$T_{\max}$ & Maximum TBT latency \\
\hline
\noalign{\vskip 2pt}
$E_k^{\left({\mathrm{dev}}\right)}$ & Total energy cost on device \\[4pt]
$E_{\max}^{\left({\mathrm{dev}}\right)}$ & Maximum device energy budget \\[4pt]
$T_{k, l}^{\left({\mathrm{comp}}\right)}$ & Computation latency \\[4pt]
$R_k^{\left({\mathrm{ul}}\right)}, R_k^{\left({\mathrm{dl}}\right)}$ & Communication rate\\[4pt]
$B_k^{\left({\mathrm{ul}}\right)}, B_k^{\left({\mathrm{dl}}\right)}$ & Communication bandwidth\\[4pt]
$P_k^{\left({\mathrm{ul}}\right)}, P_k^{\left({\mathrm{dl}}\right)}$ & Communication power\\[4pt]
$E_{k}^{\left({\mathrm{ul}}\right)}, E_{k}^{\left({\mathrm{dl}}\right)}$ & Communication energy\\[4pt]
$T_{k, l}^{\left({\mathrm{ul}}\right)}, T_{k, l}^{\left({\mathrm{dl}}\right)}$ & Communication latency in one iteration\\[4pt]
\hline
\end{tabular}
\end{table}

\section{Related Work}\label{sec: related_work}
In this section, we introduce several related works, including LLM inference-time scaling, wireless network-aided LLMs, and MoE-aided DRL.

\subsection{Inference-Time Scaling in LLMs}
Inference-time scaling~\cite{google-inference-scaling} introduces additional computation during inference to overcome the single-pass reasoning bottleneck without increasing model parameters. Existing methods can be categorized into explicit and implicit reasoning. Explicit approaches, such as CoT prompting and its variants~\cite{cot-2022,zero-shot-reasoner}, generate intermediate reasoning steps that are often optimized with reinforcement learning to improve interpretability and accuracy. For example, zero-shot CoT~\cite{zero-shot-reasoner} employs a simple prompt phrase \emph{Let's think step by step} to elicit multi-step reasoning in LLMs. However, these methods introduce redundant tokens and often struggle with tasks that lack clearly decomposable substeps \cite{jin2025reasoningornot}. To overcome these limitations, implicit reasoning shifts the reasoning process into the model's latent space to avoid additional token-level outputs. One major line of work in this direction is knowledge distillation, where reasoning ability is transferred from large to small models through intermediate representations~\cite{distill-step-by-step,implicit-cot}. For example, multi-task distillation~\cite{distill-step-by-step} trains the student to jointly learn answers and intermediate rationales from the teacher, while implicit CoT distills the hidden CoT representations into the student~\cite{implicit-cot}. These approaches enable implicit reasoning that outputs only the final prediction during inference, thereby reducing token-level overhead but at the cost of generalization across tasks. Another emerging direction explores latent-space reasoning, where models operate on hidden representations instead of generating explicit CoT tokens. 
Some methods such as \emph{continuous thought}~\cite{coconut} feed hidden states back into the model for multi-step reasoning, while recurrent latent reasoning models~\cite{geiping2025scaling} scale computation by iteratively refining these hidden states. These approaches achieve strong reasoning performance without requiring extra tokens or specialized training data.
While these methods address the limitations of explicit schemes and improve task generalization, they still lack flexible control over reasoning computation. These limitations motivate more adaptive and resource-aware frameworks that can dynamically balance reasoning performance and computational efficiency.

\subsection{Wireless Network-aided LLMs}
Wireless network-aided LLMs leverage cloud–edge collaboration to offload computation from user devices to nearby Mobile Edge Computing (MEC) nodes or the cloud, thereby enhancing their feasibility in resource-constrained environments~\cite{MECsurvey}.
Existing studies can be categorized into two directions, including wireless network-aided training and inference. For LLM training, two main approaches are employed, namely Federated Learning (FL), which enables privacy-preserving collaboration across distributed devices for data-limited scenarios~\cite{FedLLM,OpenFedLLM}, and Split Learning (SL), which partitions model execution between devices and servers to address compute-limited scenarios~\cite{SplitLLM,SplitLoRA}. Building on both, FedsLLM~\cite{FedSplitLLM} unifies these paradigms with low-rank adaptation to enhance communication efficiency and scalability in wireless training.
Collectively, these approaches make distributed LLM training practical for wireless environments.
Although efficient training lays the foundation for LLM development, real-world deployment is dominated by the inference stage, which handles massive user queries and ultimately determines overall efficiency and responsiveness.
In wireless network-aided inference, existing approaches primarily address two key bottlenecks, including computation and communication constraints.
For computation-constrained scenarios, split inference partitions models between devices and servers to mitigate device-side computation and energy burdens.
For instance,~\cite{EdgeShard} optimizes partition points for efficiency, while~\cite{CE-CoLLM} dynamically adapts split inference to device capabilities and network conditions, thereby alleviating device workload and maintaining stable latency.
For communication-constrained scenarios, model compression techniques such as pruning and quantization are employed to compress intermediate features of LLMs, thereby reducing bandwidth consumption and improving end-to-end latency.
Specifically,~\cite{LLMQuant} jointly optimizes layer placement and quantization precision, while~\cite{LLMPrune} applies structured pruning to reduce communication latency.
These compression strategies significantly improve data transfer efficiency to ease communication burdens.
In summary, wireless network-aided inference effectively mitigates both computation and communication bottlenecks, enabling scalable LLM deployment across dynamic wireless environments.

\subsection{MoE-aided DRL}
MoE architectures employ a gating network that activates only the most relevant experts and have been extensively explored for network optimization in MEC.
Existing MoE studies generally fall into two types, namely channel-aware and task-aware approaches. While channel-aware MoE leverages wireless channel state information to route data through experts that are best adapted to time-varying channel conditions~\cite{song2025mixture}, task-aware MoE assigns diverse computation subtasks to specialized experts, improving adaptability and efficiency in heterogeneous MEC environments~\cite{li2025theory,xue2025wdmoe}.
In parallel, DRL has also been widely applied to MEC network optimization, where agents learn to maximize system performance under complex and dynamic conditions~\cite{bi2021lyapunov}. Existing DRL-based methods are typically grouped into computation-oriented and communication-oriented policies.
Specifically, computation-oriented policies dynamically decide whether to execute tasks locally, remotely, or collaboratively. They also allocate Central Processing Unit (CPU) and Graphics Processing Unit (GPU) resources to balance latency and energy consumption~\cite{bi2021lyapunov}.
Conversely, communication-oriented policies jointly manage available radio resources to satisfy diverse QoS requirements in a wireless fading and interference environment~\cite{yan2020offloading}.
To harness the complementary strengths of conditional computation and policy learning, recent studies have integrated MoE layers into DRL architectures, with experts specialized for either communication-oriented or computation-oriented scenarios.
In communication tasks such as spectrum allocation, power control, and user scheduling, MoE provides fine-grained adaptability to dynamic channel conditions while DRL agents learn global coordination strategies~\cite{liu2024meta, du2024mixture}.
Similarly, in computation-oriented scenarios, this integrated framework facilitates efficient edge–cloud offloading and resource scheduling by managing diverse workloads under DRL-driven decision-making~\cite{willi2024mixture, wang2024toward}.
However, most studies address communication and computation separately, limiting adaptability across heterogeneous tasks. This motivates the integration of task-aware expert design with joint computation–communication resource optimization.

\section{System Model}\label{sec: system_model}
In this section, we introduce the latent LLM reasoning and its distributed architecture as adopted in the MORES framework with the computation and communication system models. We further formulate the resource allocation problem that naturally arises within this framework.

\begin{figure}[t]
\centering
\includegraphics[width=0.9\columnwidth]{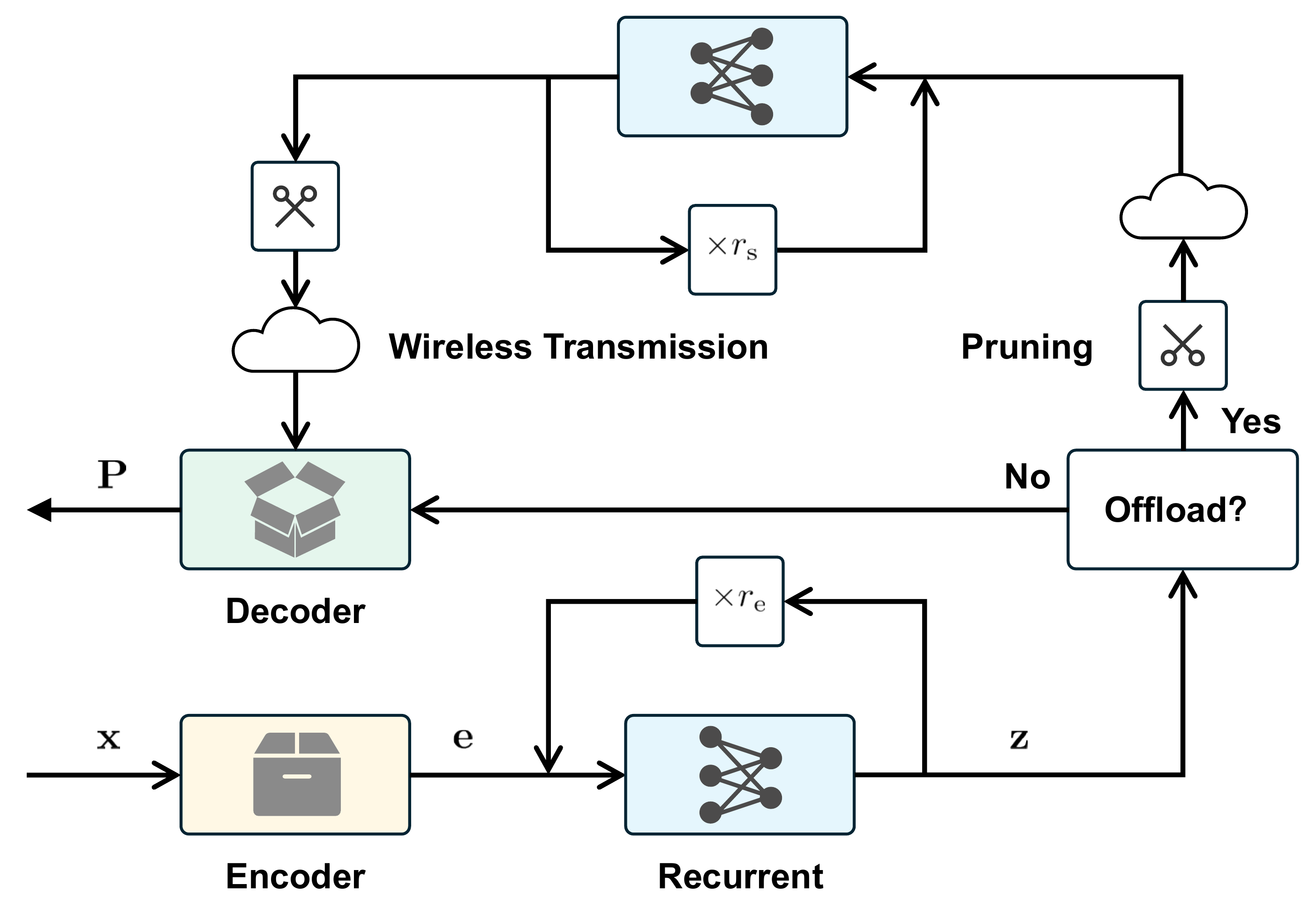}
\caption{System model for distributed latent LLM reasoning. The prelude encoder and coda decoder are executed on the device, while recurrent reasoning units are deployed across both the device and the server, enabling collaborative reasoning over a wireless channel.}
\label{fig: system_model}
\end{figure}

\subsection{Latent LLM Reasoning}
LLM reasoning involves performing multi-step inference to transform input information into a final output through intermediate computational steps.
During these computational steps, the LLM processes and refines its understanding of the input to generate output tokens that align with the intended response.
However, traditional LLM reasoning typically requires a full forward pass for each reasoning step, resulting in high computational and memory overhead.
Latent LLM reasoning mitigates this issue by decoupling the reasoning process from the full network evaluation.
Instead of processing every layer, the model projects the input into a latent space and iteratively refines a hidden state using a lightweight recurrent unit.
This approach preserves the dynamics of step-by-step reasoning while significantly reducing computation.
It mirrors the iterative nature of human thinking, where ideas are refined in memory before being verbalized.

We formalize the process as follows. Let $\mathbf{x} \in \mathcal{V}^{n}$ denote a sequence of $n$ input tokens from the vocabulary set $\mathcal{V}$. 
The input is first embedded and passed through a prelude encoder $\mathcal{P}$ to produce a latent representation $\mathbf{e}$ as
\begin{equation}\label{eqn:prelude_operation}
\mathbf{e} = \mathcal{P}(\mathbf{x}), \quad \mathbf{e} \in \mathbb{R}^{n \times v},
\end{equation}
where $v$ is the hidden size of the latent space. Latent reasoning then proceeds by recurrent updates within this latent space. It starts from an initial latent state $\mathbf{z}_{0}$ sampled from a learned noise distribution or a Gaussian prior $\mathcal{N}(0, \sigma_z^2 \cdot \mathbf{I})$, where $\sigma_z$ denotes the latent state noise scale and $\mathbf{I}$ is the identity matrix. A recurrent reasoning unit $\mathcal{R}$ then performs iterative updates as
\begin{equation}\label{eqnrecurrent_operation}
\mathbf{z}_{i} = \mathcal{R}(\mathbf{e}, \mathbf{z}_{i-1}), \quad i = 1, 2, \dots, r,
\end{equation}
where $r$ denotes the number of recurrent steps, and each $\mathbf{z}_{i} \in \mathbb{R}^{v}$ summarizes the internal state of the model after the $i$-th latent update. Finally, a coda decoder $\mathcal{C}$ maps the final latent state $\mathbf{z}_{r}$ to a probability distribution over output tokens as
\begin{equation}\label{eqncoda_operation}
\mathbf{P} = \mathcal{C}(\mathbf{z}_{r}), \quad \mathbf{P} \in [0, 1]^{|\mathcal{V}|},
\end{equation}
where $\mathbf{P}$ represents the softmax scores for decoding.

\begin{figure}[t]
\centering
\includegraphics[width=1.0\columnwidth]{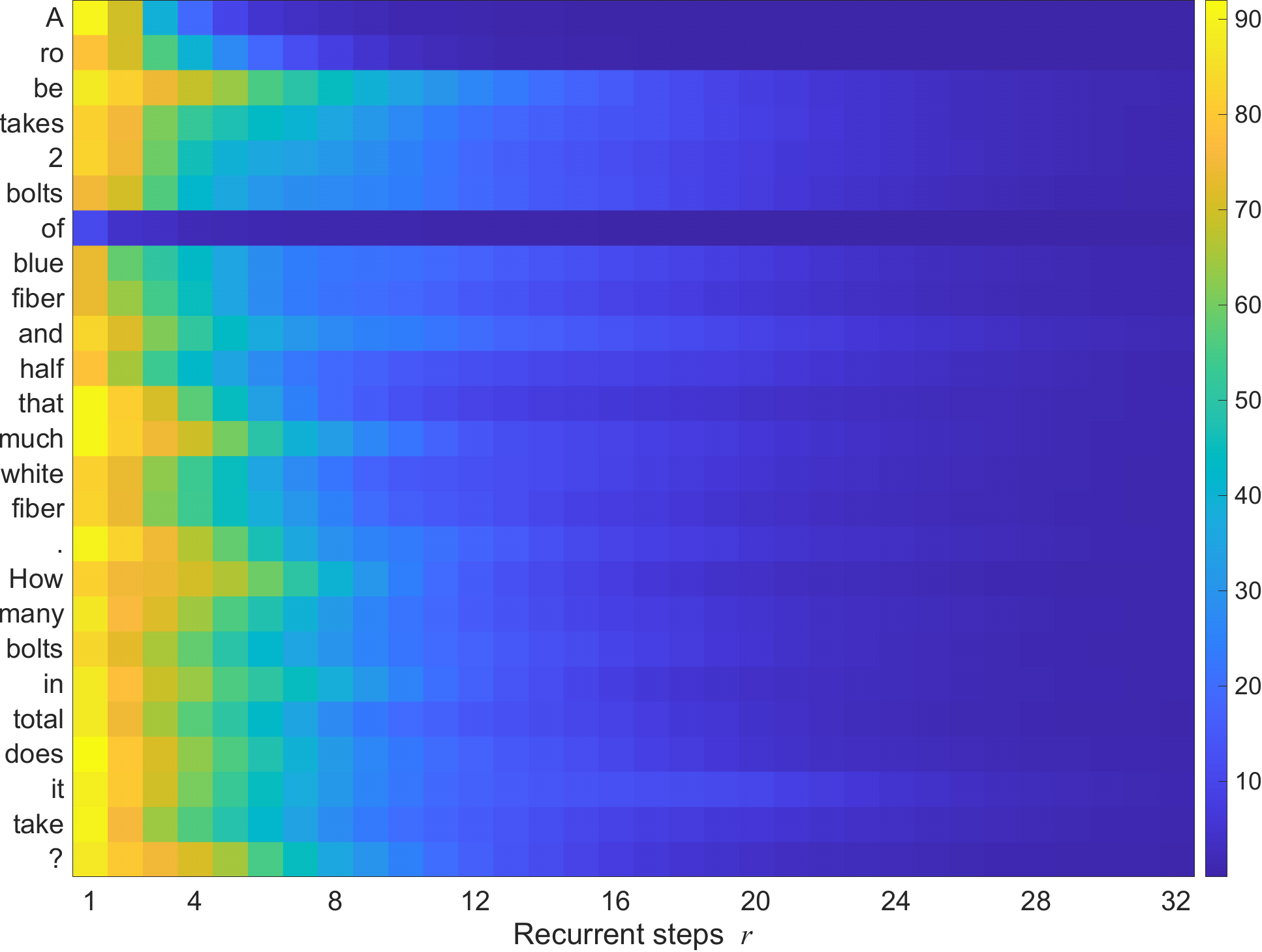}
\caption{Token-wise reasoning depth required for full convergence in latent space.}
\label{fig: heatmap}
\end{figure}

\subsection{Distributed Reasoning}
Traditional LLM reasoning assumes the entire model resides on a single high-performance node, e.g., an A100 GPU, which imposes strict hardware demands and limits deployment flexibility. To enable LLM reasoning on resource-constrained devices, we propose a distributed reasoning framework that offloads several reasoning steps to an edge server while retaining lightweight processing locally. The proposed system model is illustrated in Fig.~\ref{fig: system_model}. This design allows low-power devices to handle input encoding and preliminary reasoning while delegating deeper inference to centralized infrastructure.

As shown in Fig.~\ref{fig: heatmap}, latent reasoning may exhibit uneven convergence patterns across tokens. Some tokens may converge quickly, while others require deeper iterative refinement. This observation motivates a collaborative setting, where the device performs initial encoding and the server extends the reasoning depth when needed.

In MORES, the prelude encoder $\mathcal{P}$ is deployed on the edge device to transform the prompt input $\mathbf{x}$ into a latent representation $\mathbf{e}$. A lightweight recurrent reasoning unit $\mathcal{R}$ then performs the initial $r_{\text{e}}$ steps of iterative reasoning locally on the edge device, updating the latent state from $\mathbf{z}_{0}$ to $\mathbf{z}_{r_{\text{e}}}$. Both the latent representation $\mathbf{e}$ and the intermediate state $\mathbf{z}_{r_{\text{e}}}$ are then transmitted to the server, where deeper inference continues for the remaining ${r} - r_{\text{e}}$ steps. The final latent state $\mathbf{z}_{r}$ is sent back to the edge device, where the coda decoder $\mathcal{C}$ maps it to token probabilities for generation.
Note that $r_{\text{e}}$ is an environmental parameter determined by the capabilities of the edge device. In the MORES framework, the server acts as a supplementary service to assist in completing the reasoning process. Consequently, setting $r_{\text{e}} = {r}$ corresponds to purely local reasoning, while $r_{\text{e}} = 0$ represents offloading the entire recurrent reasoning to the server.
The complete procedure is summarized in Algorithm~\ref{algdistributed_reasoning}.

\begin{algorithm}[t]
\caption{Distributed Latent LLM Reasoning for One Token Generation}
\label{algdistributed_reasoning}
\begin{algorithmic}[1]
\State \textbf{Input:} Input request $\mathbf{x}_k$, total recurrent steps $r$, edge recurrent steps $r_{\text{e}}$
\State \textbf{Output:} Generated one token for the prompt input $\mathbf{x}_k$ 
\vspace{0.3em}
\Statex \textbf{[Edge Device]}
\State Encode prompt: $\mathbf{e}_k \leftarrow \mathcal{P}(\mathbf{x}_k)$
\State Initialize latent state: $\mathbf{z}_{0} \sim \mathcal{N}(0, \sigma_z^2 \cdot \mathbf{I})$
\For{$i = 1$ to $r_{\text{e}}$}
    \State $\mathbf{z}_{i} \leftarrow \mathcal{R}(\mathbf{e}_k, \mathbf{z}_{i-1})$
\EndFor
\State Transmit $(\mathbf{e}_k, \mathbf{z}_{r_{\text{e}}})$ to server
\vspace{0.3em}
\State  $\#$ Uplink Wireless Transmission
\vspace{0.3em}
\Statex \textbf{[Edge Server]}
\For{$i = r_{\text{e}} + 1$ to $r$}
    \State $\mathbf{z}_{i} \leftarrow \mathcal{R}(\mathbf{e}_k, \mathbf{z}_{i-1})$
\EndFor
\State Transmit $\mathbf{z}_{r}$ to device
\vspace{0.3em}
\State  $\#$ Downlink Wireless Transmission
\vspace{0.3em}
\Statex \textbf{[Edge Device]}
\State Decode token: $\mathbf{P}_k \leftarrow \mathcal{C}(\mathbf{z}_{r})$
\State \Return $\arg\max\limits_{\mathcal{V}}\mathbf{P}_k$
\end{algorithmic}
\end{algorithm}

\subsection{Computation Model}
In this section, we present the computation model of the MORES
framework. A sequence of $K$ reasoning requests arrives over time, indexed by $\mathcal{K} = \{1, 2,\dots, K\}$. Each request $k \in \mathcal{K}$ is served in a corresponding time slot $t_k$.
The task set is denoted as $\mathcal{D} = \{1, 2,\dots, D\}$, where the task type of request $k$ is $d_k \in \mathcal{D}$.
Specifically, each request $k$ consists of an input prompt $\mathbf{x}_k$ with $n_k$ input tokens and produces an output sequence $\mathbf{y}_k$ with $m_k$ tokens. Output tokens are generated autoregressively. The LLM first processes $\mathbf{x}_k$ to produce the first output token and subsequently generates each additional token based on the previous outputs. This results in $m_k$ rounds of computation and communication.

Let each generation round be indexed by $l \in \{1, 2, \dots, m_k \}$. The number of tokens processed in the $l$-${\rm th}$ round is denoted by $\alpha_{k, l}$, where
\begin{equation}\label{eqn: num_token_in_first_round}
\alpha_{k, 1} = n_k, \quad \forall k,
\end{equation}
and
\begin{equation}\label{eqn: num_token_in_rest_round}
\alpha_{k, l} = 1, \quad \forall l \neq 1, \forall k.
\end{equation}

In each round $l$ for request $k$, the device incurs a computational energy cost of $\alpha_{k,l} \cdot \left(E_p + r_{\text{e}} \cdot E_r + E_c\right)$, where $E_p$, $E_r$, and $E_c$ denote the per-token energy costs of encoding, recurrence, and decoding, respectively. 
The server performs the remaining $r_k - r_{\text{e}}$ reasoning steps, resulting in a recurrence cost of $\alpha_{k,l} \cdot \left({r_k - r_{\text{e}}}\right)$.
The latency of the computation model is proportional to the recurrent steps $r_k$ and the number of tokens $\alpha_{k, l}$~\cite{yu2022orca}, given by
\begin{equation}\label{eqn: comp_latency}
T_{k, l}^{\left({\mathrm{comp}}\right)} = c_1 \cdot r_k \cdot \alpha_{k, l} + c_2,
\end{equation}
where $c_1$ and $c_2$ are two constants determined by the underlying hardware and the employed model.

\subsection{Communication Model}
Under the MORES framework, each request involves two communication stages, comprising the uplink transmission of both the latent representation $\mathbf{e}_k$ and the latent state $\mathbf{z}_{r_{\text{e}}}$ from the edge device to the server, and the downlink transmission of the final latent state $\mathbf{z}_{r}$ from the server back to the device.
We consider a block fading wireless channel with a path loss factor $\eta$, where the uplink and downlink channel gains, denoted by $h_k^{\left({\mathrm{ul}}\right)}$ and $h_k^{\left({\mathrm{dl}}\right)}$, remain constant during the processing of one request. To simplify the problem, we assume $h_k=h_k^{\left({\mathrm{ul}}\right)}=h_k^{\left({\mathrm{dl}}\right)}$ without loss of generality. The channel gain $h_k$ is assumed to be unknown to both the device and the server.

The uplink and downlink data rates $R_k^{\left({\mathrm{ul}}\right)}$ and $R_k^{\left({\mathrm{dl}}\right)}$ for the $k$-${\rm th}$ request are given by
\begin{equation}\label{eqn: uplink_shannon_rate}
R_k^{\left({\mathrm{ul}}\right)} = B_k^{\left({\mathrm{ul}}\right)} \cdot \log\left({1 + \eta \cdot h_k^2 \cdot \frac{P_k^{\left({\mathrm{ul}}\right)}}{\sigma_h^2}}\right),
\end{equation}
\begin{equation}\label{eqn: downlink_shannon_rate}
R_k^{\left({\mathrm{dl}}\right)} = B_k^{\left({\mathrm{dl}}\right)} \cdot \log\left({1 + \eta \cdot h_k^2 \cdot \frac{P_k^{\left({\mathrm{dl}}\right)}}{\sigma_h^2}}\right),
\end{equation}
where $B_k^{\left({\mathrm{ul}}\right)}$, $B_k^{\left({\mathrm{dl}}\right)}$, $P_k^{\left({\mathrm{ul}}\right)}$, and $P_k^{\left({\mathrm{dl}}\right)}$ represent the bandwidths and transmitted powers in the uplink and downlink transmissions, respectively, and $\sigma_h^2$ is the noise power. For analytical simplicity, we consider uniform power and bandwidth allocation across requests as
\begin{equation}\label{eqn: unified_comm_power}
P_k^{\left({\mathrm{ul}}\right)} = P^{\left({\mathrm{ul}}\right)}, \quad  P_k^{\left({\mathrm{dl}}\right)} = P^{\left({\mathrm{dl}}\right)}, \quad \forall k,
\end{equation}
\begin{equation}\label{eqn: unified_comm_bandwidth}
B_k^{\left({\mathrm{ul}}\right)} = B^{\left({\mathrm{ul}}\right)}, \quad  B_k^{\left({\mathrm{dl}}\right)} = B^{\left({\mathrm{dl}}\right)}, \quad \forall k.
\end{equation}

To reduce uplink overhead, we introduce a pruning rate $\rho_k \in [0, 1)$, which denotes the fraction of latent information pruned before transmission. We assume full transmission on the downlink, where $\rho_k^{\left(\mathrm{dl}\right)} = 0$. Let $q$ denote the quantization bit width for each real-valued element, and recall that $v$ denotes the hidden size. Thus, each token requires $q \cdot v$ bits for transmission.

For request $k$ in round $l$, the uplink and downlink communication latencies $T_{k, l}^{\left({\mathrm{ul}}\right)}$ and $T_{k, l}^{\left({\mathrm{dl}}\right)}$ are derived as
\begin{equation}\label{eqn: single_round_comm_latency}
T_{k, l}^{\left({\mathrm{ul}}\right)} = 2\left({1-\rho_k}\right) \cdot \frac{q \cdot v \cdot \alpha_{k, l}}{R_k^{\left({\mathrm{ul}}\right)}}, \quad T_{k, l}^{\left({\mathrm{dl}}\right)} = \frac{q \cdot v \cdot \alpha_{k, l}}{R_k^{\left({\mathrm{dl}}\right)}}.
\end{equation}

The corresponding energy consumptions are
\begin{equation}\label{eqn: single_round_comm_energy}
E_{k, l}^{\left({\mathrm{ul}}\right)} = P_k^{\left({\mathrm{ul}}\right)} \cdot T_{k, l}^{\left({\mathrm{ul}}\right)}, \quad E_{k, l}^{\left({\mathrm{dl}}\right)} = P_k^{\left({\mathrm{dl}}\right)} \cdot T_{k, l}^{\left({\mathrm{dl}}\right)}.
\end{equation}

Summing across all $m_k$ rounds of token generation, the total uplink and downlink energies for the $k$-${\rm th}$ request are
\begin{equation}\label{eqn: total_uplink_comm_energy}
\begin{aligned}
E_{k}^{\left({\mathrm{ul}}\right)} =   \frac{2\left({1-\rho_k}\right) \cdot q \cdot v \cdot \left({n_k + m_k - 1}\right) \cdot P^{\left({\mathrm{ul}}\right)}}{R_k^{\left({\mathrm{ul}}\right)}},
\end{aligned}
\end{equation}
\begin{equation}\label{eqn: total_downlink_comm_energy}
\begin{aligned}
E_{k}^{\left({\mathrm{dl}}\right)} =  \frac{q \cdot v \cdot \left({n_k + m_k - 1}\right) \cdot P^{\left({\mathrm{dl}}\right)}}{R_k^{\left({\mathrm{dl}}\right)}}.
\end{aligned}
\end{equation}

These equations capture the total communication cost incurred when serving the $k$-${\rm th}$ request, reflecting the uplink and downlink transfers of latent intermediate features under optional pruning.

\subsection{Problem Formulation}
In this section, we model the resource allocation problem that arises under the MORES framework, considering both computation and communication costs in wireless networks. Let $\delta_k \in \{0, 1\}$ denote the correctness of the reasoning result for the $k$-${\rm th}$ request, where $\delta_k = 1$ denotes a correct LLM answer and $\delta_k = 0$ otherwise. The average system throughput over $K$ requests is denoted as
\begin{equation}
U = \frac{1}{K} \sum_{k=1}^K \delta_k.
\end{equation}

The total energy consumption on the device side for the $k$-${\rm th}$ request is denoted by $E_k^{\left({\mathrm{dev}}\right)}$, which can be derived as
\begin{equation}\label{eqn: device_energy}
\begin{aligned}
E_k^{\left({\mathrm{dev}}\right)} &= E_k^{\left(\mathrm{ul}\right)} + \sum_{l=1}^{m_k} \alpha_{k, l} \cdot \left({E_p + r_{\text{e}} \! \cdot \! E_r + E_c}\right)\\
&= E_k^{\left({\mathrm{ul}}\right)} + \left({n_k + m_k - 1}\right) \cdot \left({E_p + r_{\text{e}} \! \cdot \! E_r + E_c}\right).
\end{aligned}
\end{equation}

From an energy-efficient perspective, the device-side energy constraint is considered as
\begin{equation}\label{eqn: energy_constraints}
\sum_k E_k^{\left({\mathrm{dev}}\right)} \leq E_{\max}^{\left({\mathrm{dev}}\right)},
\end{equation}
where $E_{\max}^{\left({\mathrm{dev}}\right)}$ denotes the energy budget for the device.

Considering the reasoning capacity of the server, the total recurrence constraint is imposed to avoid system saturation and guarantee service stability, which is derived as
\begin{equation}\label{eqn: token_constraints}
\sum_{k, l} \alpha_{k, l} \cdot \left({r_k - r_{\text{e}}}\right) \leq G_{\max},
\end{equation}
where $G_{\max}$ represents the maximum recurrence budget for the total $K$ requests.

For each request $k$, the Time-Between-Tokens (TBT) latency constraint is introduced to guarantee a smooth user experience as
\begin{equation}\label{eqn: tbt_constraints}
T_{k, l}^{\left({\mathrm{ul}}\right)} + T_{k, l}^{\left({\mathrm{comp}}\right)} + T_{k, l}^{\left({\mathrm{dl}}\right)} \le T_{\max},
\end{equation}
where $T_{\max}$ represents the maximum TBT latency for each request. The system aims to maximize the average throughput $U$, subject to the energy and latency constraints for both computation and communication, and the reasoning capacity limit as
\begin{equation*}
\begin{alignedat}{3}
\mathbf{(P1)}\quad & \underset{\{r_k,\rho_k\}}{\max} \quad
& & U=\frac{1}{K}\sum_{k}\delta_k, \\
& \hspace{0.6em}\text{s.t.},\quad
& & \sum_k E_k^{(\mathrm{dev})} \le E_{\max}^{(\mathrm{dev})}, \\
& 
& & \sum_{k,l} \alpha_{k,l} \cdot \left({r_k - r_{\text{e}}}\right) \le G_{\max}, \\
& 
& & T_{k,l}^{(\mathrm{ul})}+T_{k,l}^{(\mathrm{comp})}+T_{k,l}^{(\mathrm{dl})}\le T_{\max},\quad \forall k,\forall l.\\
\end{alignedat}
\end{equation*}

This formulation captures the trade-offs between reasoning quality, energy efficiency, and resource availability.
In (P1), $r_k$ serves as one of the decision variables for the total recurrent steps, with the cloud providing a supplementary service for any processing beyond the fixed $r_{\text{e}}$.
Increasing the number of recurrent steps $r_k$ typically improves the correctness indicator $\delta_k$ but also increases computational latency and token-level inference overhead. On the other hand, reducing the uplink pruning rate $\rho_k$ enhances latent information quality but incurs higher transmission latency and energy consumption. The optimizer must coordinate the recurrent steps and the pruning rate across requests to balance reasoning accuracy and overall system efficiency. In addition, the TBT latency constraint introduces a specific trade-off between computation and communication latency, since increasing the recurrent steps extends computation time, thereby limiting the time available for communication. To satisfy the latency limit under fixed transmission power, the system must adopt a higher pruning rate, which in turn reduces accuracy. Hence, the optimizer must carefully adjust the recurrent steps and the pruning rate to satisfy latency requirements while preserving model performance.

\section{Semantic MoE-based DRL Algorithm}\label{sec: semantic_moe_drl}
In this section, we present our proposed semantic MoE-based DRL algorithm and explain the design of each key component, focusing on the semantic router for task-type classification and the MoE network for expert selection and adaptive reasoning.

\subsection{Deep Reinforcement Learning Solutions}
DRL formulates sequential decision-making as an agent interacting directly with an environment $\mathcal{E}$. At each discrete time step $t$, the agent observes the current state $\mathbf{s}_t \in \mathcal{S}$ and selects an action $\mathbf{a}_t \in \mathcal{A}$ according to its policy~\cite{sutton1998reinforcement}. In response, the environment emits a reward $u_t \in \mathbb{R}$ and transitions to a new state $\mathbf{s}_{t+1}$ according to the dynamics $\mathcal{S} \times \mathcal{A} \rightarrow \mathcal{S}$.
Following the above description, we formulate problem (P1) as a Markov decision process (MDP) and specify its state space, action space, and environment reward.

\emph{State space} $\mathcal{S}$ captures the instantaneous status of the system. The state $\mathbf{s}_t$ captured at time step $t$ is defined as
\begin{equation}\label{eqn: DRL_state}
\mathbf{s}_t = \{{ t, G_{\mathrm{res}, t}, E_{\mathrm{res}, t}^{\left({\mathrm{dev}}\right)}}\},
\end{equation}
where $G_{\mathrm{res}, t}$ and $E_{\mathrm{res}, t}^{(\mathrm{dev})}$ denote the remaining recurrence budget and the remaining device energy budget at time step $t$, respectively. For notational clarity, the request index $k$ and the time slot $t$ are used interchangeably without ambiguity.

\emph{Action space} $\mathcal{A}$ defines the computation-communication resource allocation for request $k$ at time step $t$ as
\begin{equation}\label{eqn: DRL_action}
\mathbf{a}_t = \{{r_t, \rho_t}\},
\end{equation}
where $r_t$ and $\rho_t$ denote the recurrent steps and the pruning rate configured for the request served in time slot $t$, respectively. These action variables are consistent with the decision variables in problem (P1).

\emph{Environment reward} $u_t$ represents the feedback signal from the environment after executing action $\mathbf{a}_t$ in state $\mathbf{s}_t$, and is set to $u_t = \delta_t$ to reflect reasoning correctness. A higher reward indicates successful reasoning with efficient resource usage.

Given the need to balance exploitation and exploration under hard constraints, we adopt the SAC algorithm to train the agent. SAC learns a stochastic policy that maximizes the expected cumulative reward while encouraging exploration via entropy regularization~\cite{haarnoja2018soft}. It is well-suited for dynamic wireless environments with uncertain channel conditions and variable resource budgets. After offline training, the learned policy can be deployed for real-time decision-making across reasoning tasks.

The SAC algorithm mainly involves the updates of the actor and critic networks, which are described as follows. The actor is optimized by maximizing the entropy-regularized policy objective
\begin{equation}\label{SAC_actor_loss}
J_{\pi}(\phi) =
\mathbb{E}_{\mathbf{s} \sim \mathcal{B},\ \mathbf{a} \sim \pi_{\phi}(\cdot|\mathbf{s})}
\Big[
Q_{\theta}(\mathbf{s}, \mathbf{a}) - \tau \log \pi_{\phi}(\mathbf{a}|\mathbf{s})
\Big],
\end{equation}
where $\mathbf{s} \sim \mathcal{B}$ denotes sampling states from the replay buffer $\mathcal{B}$, and $\mathbf{a} \sim \pi_{\phi}(\cdot|\mathbf{s})$ denotes sampling actions from the current policy conditioned on $\mathbf{s}$. The stochastic policy $\pi_{\phi}$ is parameterized by $\phi$, and $Q_{\theta}$ is the state–action value function with parameters $\theta$. The coefficient $\tau$ controls the trade-off between maximizing the expected return and the policy entropy, thereby encouraging exploration via entropy regularization while avoiding premature convergence to deterministic suboptimal policies. The critic is updated by minimizing the squared temporal-difference error, which is given by
\begin{equation}\label{SAC_critic_loss}
J_{Q}(\theta) =
\mathbb{E}_{(\mathbf{s}, \mathbf{a}, u, \mathbf{s}') \sim \mathcal{B}}
\Big[ Q_{\theta}(\mathbf{s}, \mathbf{a}) - y(\mathbf{s}, \mathbf{a}, u, \mathbf{s}') \Big]^{2},
\end{equation}
with the soft Bellman target given by
\begin{equation}\label{SAC_soft_target}
y(\mathbf{s}, \mathbf{a}, u, \mathbf{s}') =
u + \gamma \Big[ Q_{\theta'}(\mathbf{s}', \mathbf{a}')
- \tau \log \pi_{\phi}(\mathbf{a}'|\mathbf{s}') \Big],
\end{equation}
where $\mathbf{a}' \sim \pi_{\phi}(\cdot|\mathbf{s}')$ denotes the next action sampled from the current policy.
In~\eqref{SAC_soft_target}, $\gamma$ is the discount factor and $\mathbf{s}'$ is the next state in the sampled transition. The target network $Q_{\theta'}$ is updated via soft updates to stabilize value estimation. This critic update encourages consistency between the current $Q$-value estimate and the soft Bellman target, thereby stabilizing value learning under entropy regularization.

\subsection{Semantic Router}
\begin{figure}[t]
\centering
\includegraphics[width=1.0\columnwidth]{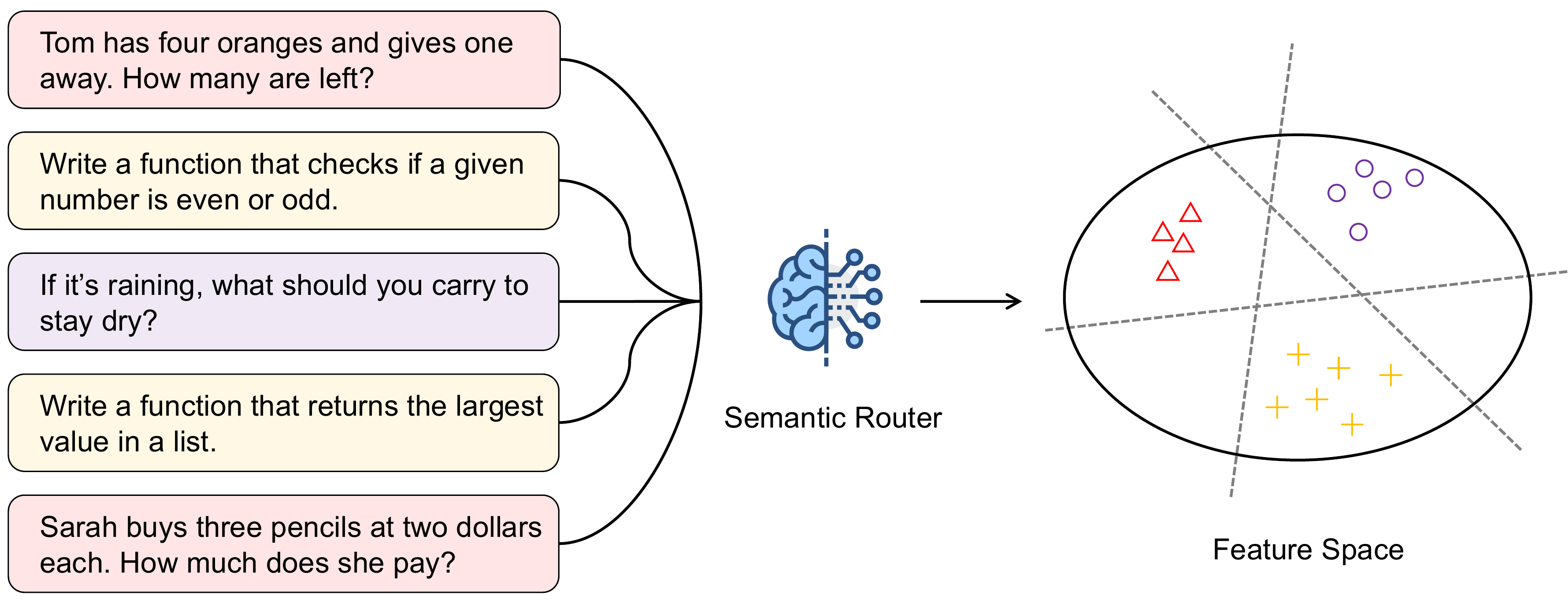}
\caption{Illustration of the semantic routing process, where the semantic router classifies different types of prompt inputs in the feature space through a trained network.}
\label{fig: semantic_router_illustration}
\end{figure}

To achieve high-fidelity task representation and precise task mapping, the MORES framework incorporates a semantic router~\cite{manias2024semantic}, which adaptively directs each request by capturing its inherent semantic characteristics.
Specifically, for each incoming request $k$, the system employs a lightweight pre-trained encoder to extract its semantic features.
The encoder maps the prompt input $\mathbf{x}_k$ to a high-dimensional semantic embedding $\mathbf{v}_k$, which serves as a comprehensive semantic representation of the request.
The semantic router then assigns the request to the most suitable category based on the semantic distance between $\mathbf{v}_k$ and the predefined class prototypes.
An illustration of this semantic routing process and the resulting classification within the feature space is depicted in Fig.~\ref{fig: semantic_router_illustration}.

Since different tasks exhibit heterogeneous characteristics in terms of energy usage and accuracy sensitivity, the task information introduced by the semantic router is essential for maximizing system throughput under constrained resources.
By exploiting task-specific information, the semantic router enables the MORES system to allocate resources more effectively, thereby enhancing both energy efficiency and inference accuracy. In MORES, the semantic router is integrated into the DRL network architecture to differentiate processing strategies across heterogeneous tasks and scenarios. 
This design facilitates optimal resource allocation under varying task requirements and provides the foundation for the subsequent MoE network design.

\subsection{Mixture of Experts Architecture}
To better handle heterogeneous tasks and diverse scenarios, a single backbone network is often insufficient to adapt to these variations. Therefore, MORES adopts a design inspired by the MoE paradigm to enhance representational capacity and reduce negative transfer. Building on conventional MoE architectures~\cite{shazeer2017outrageously}, MORES introduces multiple expert branches and a shared backbone to capture both specialized and shared knowledge. Unlike conventional soft gating that probabilistically activates experts, our design employs a deterministic gating mechanism to select the experts for each task and scenario. The selected experts are then integrated with the shared expert to produce a fused representation, which is subsequently used to generate the final decision. In the following, we present the detailed formulation of this architecture and some analysis.

\begin{figure}[t]
\centering
\includegraphics[width=0.9\columnwidth]{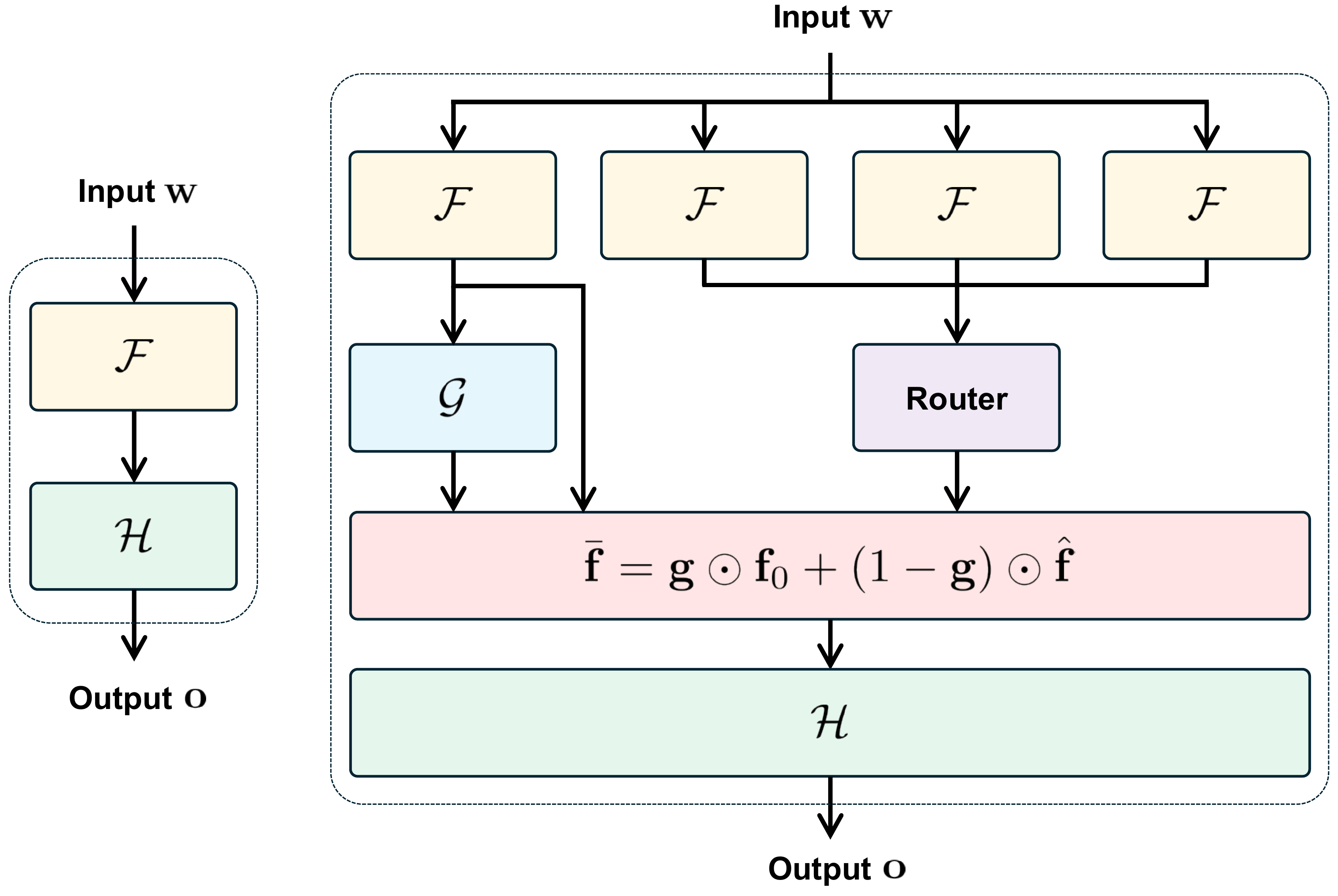}
\caption{Traditional DRL network and our proposed MoE-based network architecture design.}
\label{fig: moe_network}
\end{figure}

The traditional actor and critic networks in SAC consist of the feature extraction layer $\mathcal{F}$ and the head layer $\mathcal{H}$. First, the feature extraction layer learns the high-dimensional features from the input, and then the head layer projects the features to the final output, which can be formulated as
\begin{equation}\label{equ: basic_rl}
\mathbf{o} = \mathcal{H}\left({\mathcal{F}\left({\mathbf{w}}\right)}\right),
\end{equation}
where $\mathbf{w}$ and $\mathbf{o}$ represent the input and output, respectively. In our MoE-based architecture, $I$ parallel feature extraction networks serve as experts for executing different types of tasks, denoted as $\{\mathcal{F}_1, \mathcal{F}_2, \dots , \mathcal{F}_I \}$. Additionally, a shared feature extraction network, denoted as $\mathcal{F}_0$, is incorporated to extract the common knowledge among all types of tasks.
The architecture of our proposed MoE-based network is depicted in Fig.~\ref{fig: moe_network}.
Let $\mathbf{f}_i$ denote the features extracted by expert $\mathcal{F}_i$, where $0 \leq i \leq I$.
Through a deterministic routing mechanism conditioned on external task and scenario information, the selected features $\hat{\mathbf{f}} \in \{\mathbf{f}_1, \mathbf{f}_2, \dots , \mathbf{f}_I \}$ are merged element-wise with the common features $\mathbf{f}_0$, yielding the aggregated features $\bar{\mathbf{f}}$ as
\begin{equation}\label{equ: merged_features}
\bar{\mathbf{f}} = \mathbf{g} \odot \mathbf{f}_0 + \left({1 - \mathbf{g}}\right) \odot \hat{\mathbf{f}},
\end{equation}
where the weights are learned from the common shared features through the gating network as $\mathbf{g} = \mathcal{G}\left({\mathbf{f}_0}\right)$. The weighted features are then fed to the head layer to generate the final output as $\mathbf{o} = \mathcal{H}\left({\bar{\mathbf{f}}}\right)$.
The proposed MoE-based feature extraction is integrated into both the actor and critic networks, providing task-sensitive representations to enhance policy learning and improve value estimation across heterogeneous task and scenario distributions.

Compared with conventional MoE architectures, our design differs in three key aspects. (1) \emph{Expert selection mechanism}. In conventional MoE, the expert is chosen dynamically based on the input features provided to the gating network. In contrast, our architecture deterministically selects the expert based on task and scenario information, which provides stable and consistent routing and adapts more effectively to heterogeneous conditions. (2) \emph{Incorporation of a shared expert}. In conventional MoE, all experts are independent and parallel, without an explicit branch to capture common knowledge. To address this, our design introduces a shared expert that learns task-agnostic representations, which not only complements the task-specific experts but also enhances knowledge transfer across tasks and scenarios. (3) \emph{Fine-grained gating for feature fusion}. In conventional MoE, gating primarily serves to select and weight experts at a coarse level. However, in our design, the expert is deterministically chosen in advance, so the gating mechanism no longer selects experts but instead operates at a finer granularity by balancing the shared and task-specific features element-wise. This fine-grained feature fusion enables the model to better integrate common knowledge with task-specific representations, thereby improving generalization across heterogeneous tasks and scenarios.

This proposed architecture provides several key benefits. First, the deterministic task-dependent routing provides stable computational paths, which improves training stability and accelerates convergence in reinforcement learning. Second, by activating only the selected expert during both training and inference, the model reduces gradient interference across tasks and lowers computational overhead, leading to more efficient learning and decision-making. Third, by enabling effective integration of task-specific and shared knowledge, the architecture alleviates negative transfer and enhances generalization to heterogeneous tasks and scenarios. In addition, this modular design allows the architecture to scale flexibly by adding or removing experts as needed, enabling easy adaptation to new tasks and environments. Collectively, these benefits make our architecture particularly suitable for reinforcement learning across heterogeneous tasks and dynamic wireless environments.

Next, we analyze the computational complexity of the proposed MoE-based architecture. Let $\lvert{\mathbf{w}}\rvert$ denote the input dimension, $L$ the hidden dimension of the network, and $N$ the number of fully connected layers in the feature extraction expert. The computational cost of a single feature extraction expert is
$O\left(\lvert{\mathbf{w}}\rvert L + (N-1)L^2 + NL\right)$.
Considering both the feature extraction and output head, the overall complexity of the baseline network is 
\begin{equation}\label{equ: basic_O}
C_{\text{base}} = O\big(\lvert{\mathbf{w}}\rvert L + (N-1)L^2 + NL + \lvert{\mathbf{o}}\rvert L\big),
\end{equation}
where $\lvert{\mathbf{o}}\rvert$ denotes the output dimension. For the semantic MoE-based network featuring gating and fusion mechanisms, the overall complexity becomes
\begin{equation}\label{equ: moe_O}
C_{\text{prop}} 
= O\big(2\lvert{\mathbf{w}}\rvert L + (2N-1)L^2 + 2NL + \lvert{\mathbf{o}}\rvert L\big).
\end{equation}
When the $L^2$ term dominates the total computation, the leading terms are $(N-1)L^2$ for the baseline and $(2N-1)L^2$ for the proposed model.
This indicates that the proposed design incurs approximately twice the theoretical computation while remaining in the same asymptotic order.

Beyond computational complexity, we further examine the parameter count of the proposed architecture. The feature extraction expert contains $S=\lvert{\mathbf{w}}\rvert L + (N-1)L^2$ parameters. For a model with $I$ experts plus one shared expert, the total parameter count becomes $S \left({I + 1}\right)$, and the gating network contains $L^2$ parameters. Consequently, the proposed design introduces an additional $S I + L^2$ parameters compared with the baseline model. However, owing to sparse expert activation enabled by the semantic router, only the selected expert and the shared expert are activated during inference. Therefore, the additional parameter footprint during inference is $S + L^2$, which remains independent of the total number of experts.

\section{Experiments}\label{sec: experiments}
In this section, we present empirical insights into the MORES system to demonstrate its necessity and efficiency. The performance of the proposed semantic MoE-based SAC algorithm is then comprehensively evaluated and compared with representative baselines.

\subsection{Experimental Setup}
\textbf{Experimental Dataset.}
The MORES framework is evaluated on the GSM8K~\cite{cobbe2021gsm8k}, MBPP~\cite{austin2021program}, and HellaSwag~\cite{zellers2019hellaswag} benchmarks to address numerical, programmatic, and commonsense reasoning, respectively.
GSM8K comprises 8,792 grade-school math word problems that require sequential arithmetic operations, and has become a standard benchmark for multi-step mathematical reasoning. MBPP consists of 974 Python programming tasks with reference implementations and unit tests, providing a clear measure of code generation accuracy and functional correctness. HellaSwag contains approximately 60K multiple-choice questions derived from everyday activity descriptions, offering a rigorous test of grounded commonsense reasoning. It is worth noting that GSM8K and MBPP are generative tasks, whereas HellaSwag is formulated as a multiple-choice classification task. We built a composite dataset by sampling equally from the three datasets, ensuring that each source was fairly represented.

\textbf{Experimental Platform.} Our experiments were conducted on a workstation equipped with an NVIDIA A100 GPU (80 GB memory) and an Intel Xeon Platinum 8375C processor (32 cores). The system was running Ubuntu 20.04 LTS and employed PyTorch 2.6, together with CUDA 12.4 and cuDNN 9.1.

\textbf{Environment Details.} We train a DRL agent to allocate computation and communication resources in the MORES system with the number of requests $K=100$. Each request is sampled uniformly from the composite dataset. In addition, each request is randomly associated with a wireless channel condition, characterized by a channel gain selected uniformly from $\{0.4, 0.7, 1.2\}$, corresponding to poor, fair, and good channels, respectively. For wireless communications, the uplink and downlink bandwidths are set to 2~MHz and 20~MHz, respectively, while the uplink and downlink transmit powers are set to 0.2~W and 20~W, respectively. Without loss of generality, we set the path loss to $\eta=10^{-12}$, and the noise power density to $N_0 = -174\,\text{dBm/Hz}$. Each real-valued element is quantized to $q=16$ bits before transmission. For computation, the hidden size per token is set to $v=5280$. The per-token energy costs of encoding $E_p$ and decoding $E_c$ are both set to $2.5 \times 10^{-4}\,\text{J}$, while the per-token energy cost of the recurrent unit $E_r$ is set to $3 \times 10^{-4}\,\text{J}$.
The maximum TBT latency is set to $T_{\max} = 2.5 \times 10^{-2}$, while the computational latency parameters are set to $c_1 = 5 \times 10^{-4}$ and $c_2 = 0$.
For simplicity, the number of recurrent steps is chosen from $\{8, 12, 16, 20, 24, 28, 32\}$, and the pruning rate is selected from $\{0, 2\%, 4\%, 6\%, 8\%\}$.

\subsection{Experimental Analysis}
\begin{figure*}[t]
  \centering
  \subfloat[GSM8K token count distribution.]{%
    \includegraphics[width=0.33\textwidth]{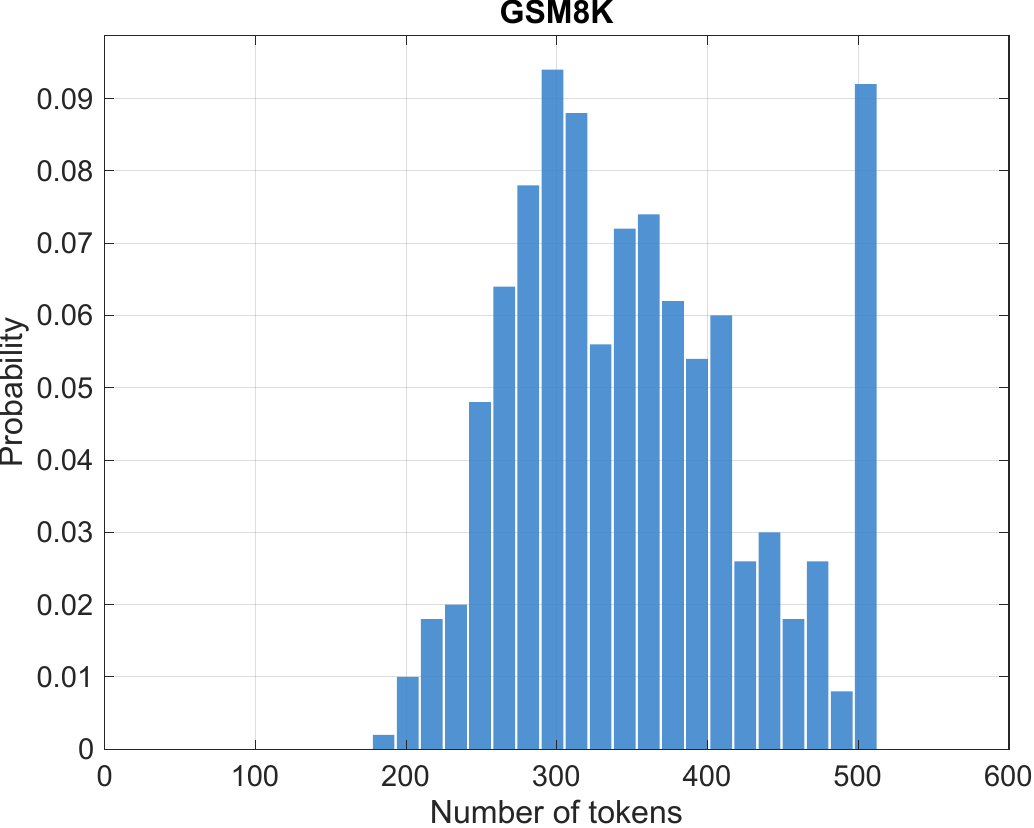}\label{fig: gsm8k_num_tokens}}
  \hfil
  \subfloat[MBPP token count distribution.]{%
    \includegraphics[width=0.33\textwidth]{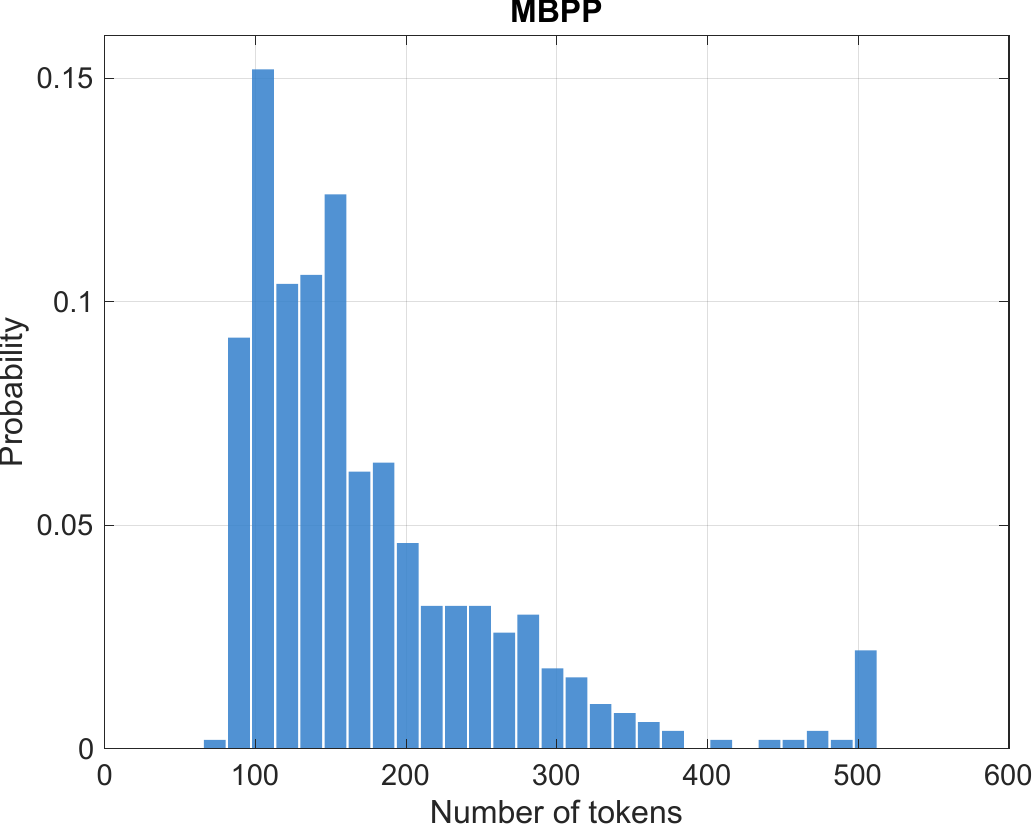}\label{fig: mbpp_num_tokens}}
  \hfil
  \subfloat[HellaSwag token count distribution.]{%
    \includegraphics[width=0.33\textwidth]{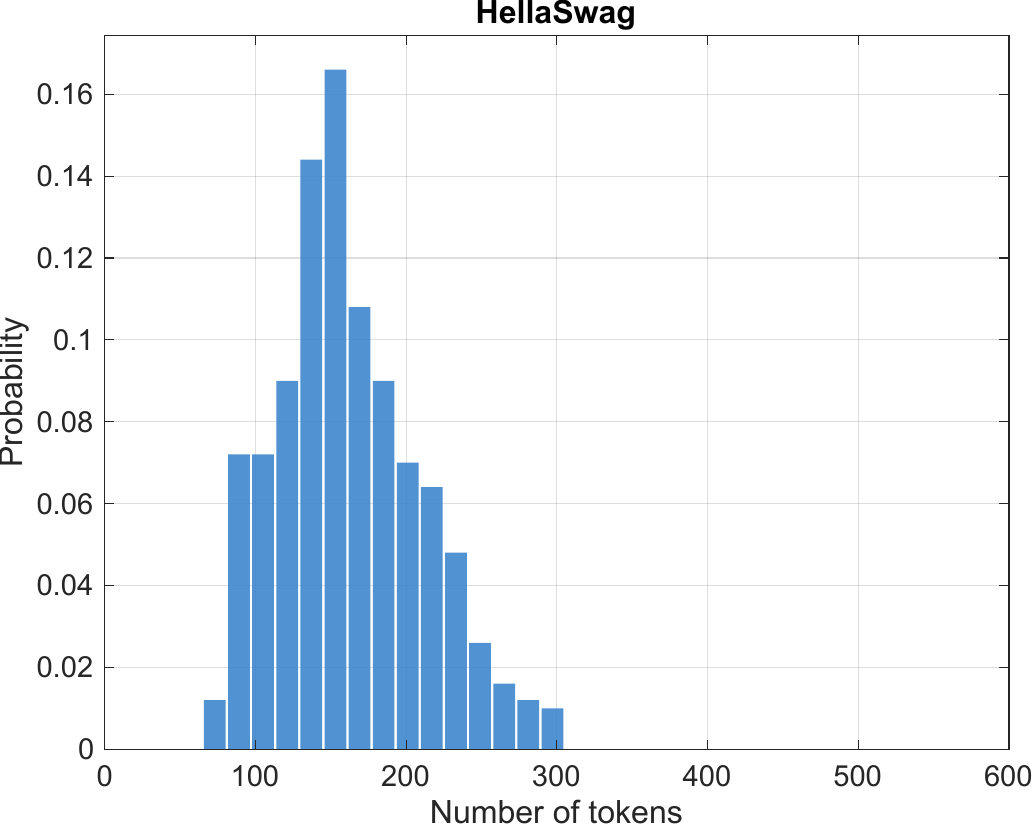}\label{fig: hella_num_tokens}}
  \caption{Histogram of token count distributions of the GSM8K, MBPP, and HellaSwag datasets.}
  \label{fig: insight_num_tokens}
\end{figure*}

Fig.~\ref{fig: insight_num_tokens} illustrates the token-length distributions of GSM8K, MBPP, and HellaSwag, demonstrating the heterogeneity of reasoning tasks in terms of sequence length. First, the three datasets exhibit different degrees of length variability. Specifically, MBPP exhibits the greatest dispersion in token length and GSM8K shows moderate dispersion, whereas HellaSwag is tightly concentrated within a narrow range. Intuitively, a narrower distribution implies more stable sequence lengths and thus simplifies scheduling decisions. In contrast, a wider distribution induces larger length fluctuations, making scheduling decisions more challenging. Second, the average token length also varies across datasets. 
GSM8K has the longest sequences on average, followed by MBPP and HellaSwag. Consequently, tasks with longer sequences generally incur higher computation and transmission energy costs during reasoning. Overall, these results confirm substantial sequence-length heterogeneity across tasks, suggesting that task-dependent scheduling is necessary to support efficient reasoning.

\begin{figure*}[!t]
  \centering
  \subfloat[GSM8K accuracy vs. pruning rate.]{%
    \includegraphics[width=0.32\textwidth]{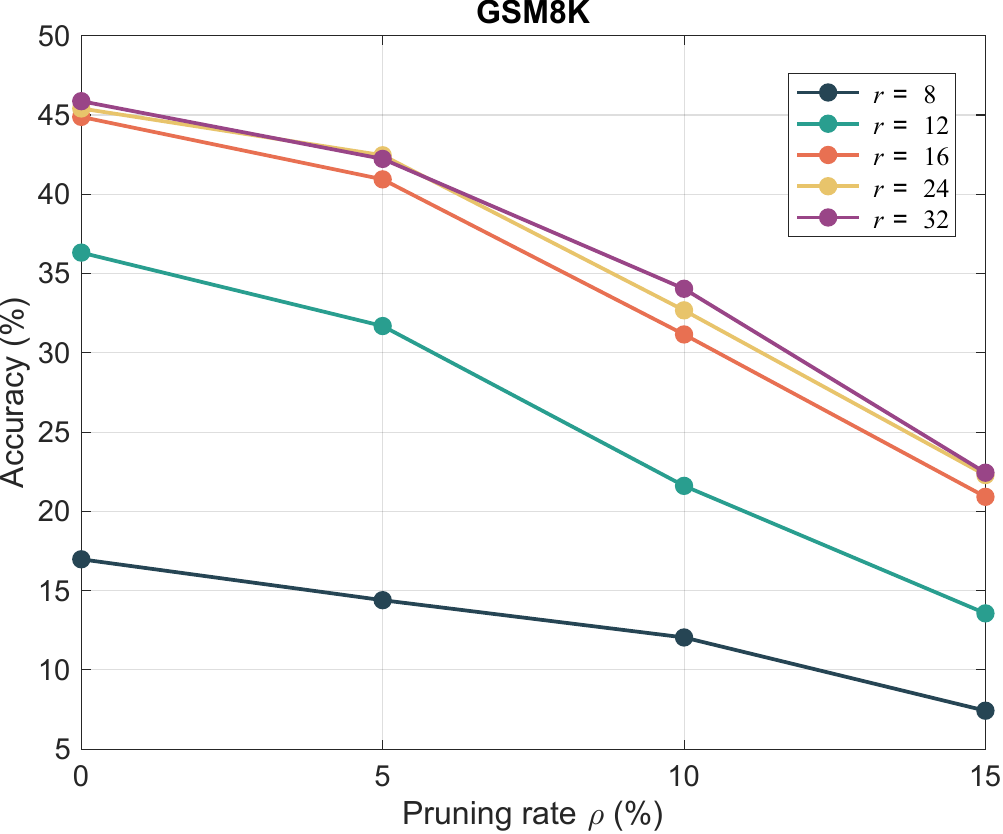}\label{fig: gsm8k_vs_pruning}}
  \hfil
  \subfloat[MBPP accuracy vs. pruning rate.]{%
    \includegraphics[width=0.32\textwidth]{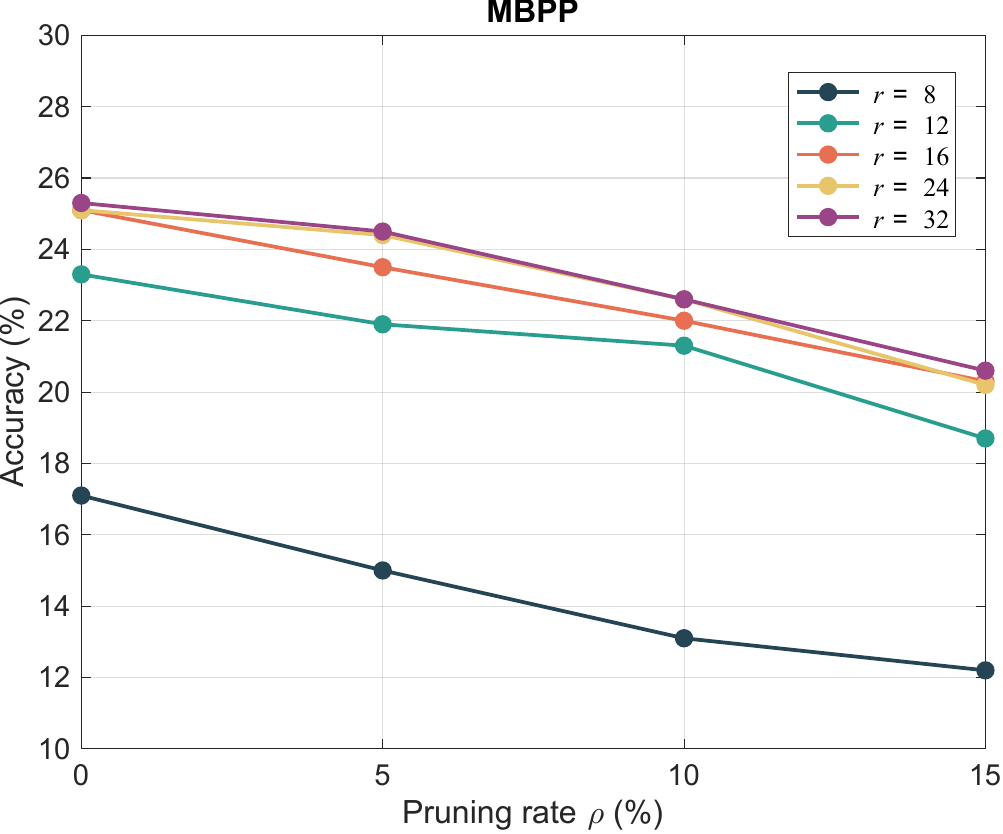}\label{fig: mbpp_vs_pruning}}
  \hfil
  \subfloat[HellaSwag accuracy vs. pruning rate.]{%
    \includegraphics[width=0.32\textwidth]{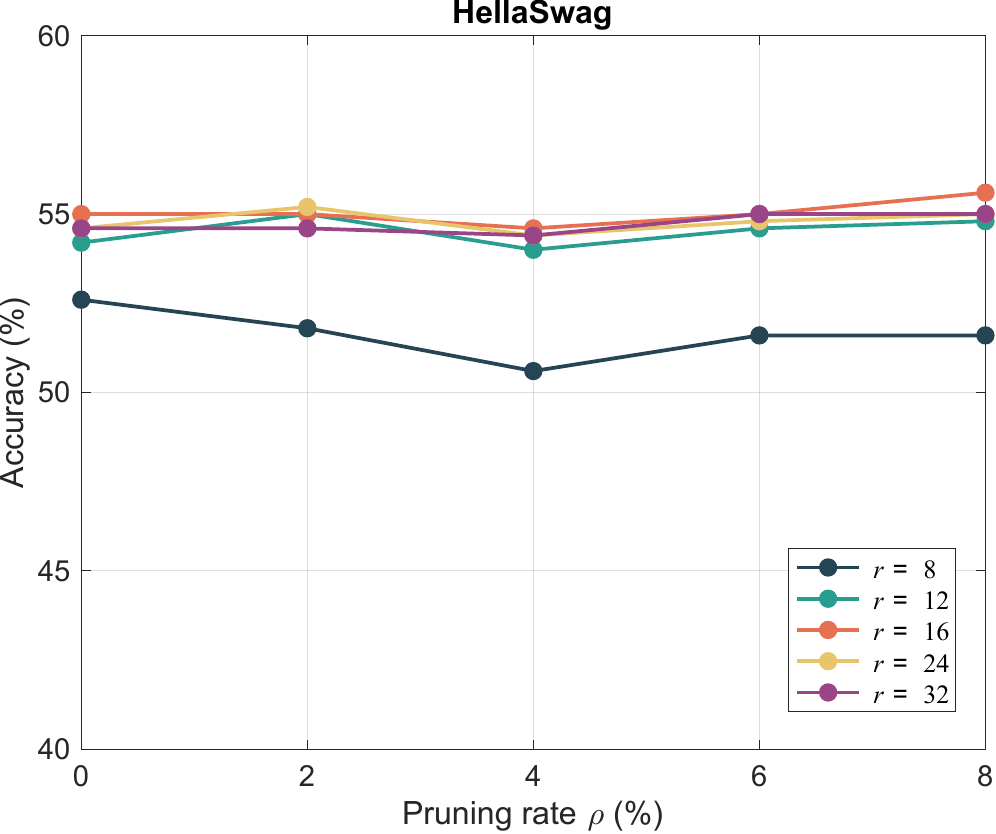}\label{fig: hella_vs_pruning}}
  \caption{Accuracy vs. pruning rate under different recurrent steps.}
  \label{fig: insight_vs_pruning}
\end{figure*}

\begin{figure*}[!t]
  \centering
  \subfloat[GSM8K accuracy vs. recurrent steps.]{%
    \includegraphics[width=0.32\textwidth]{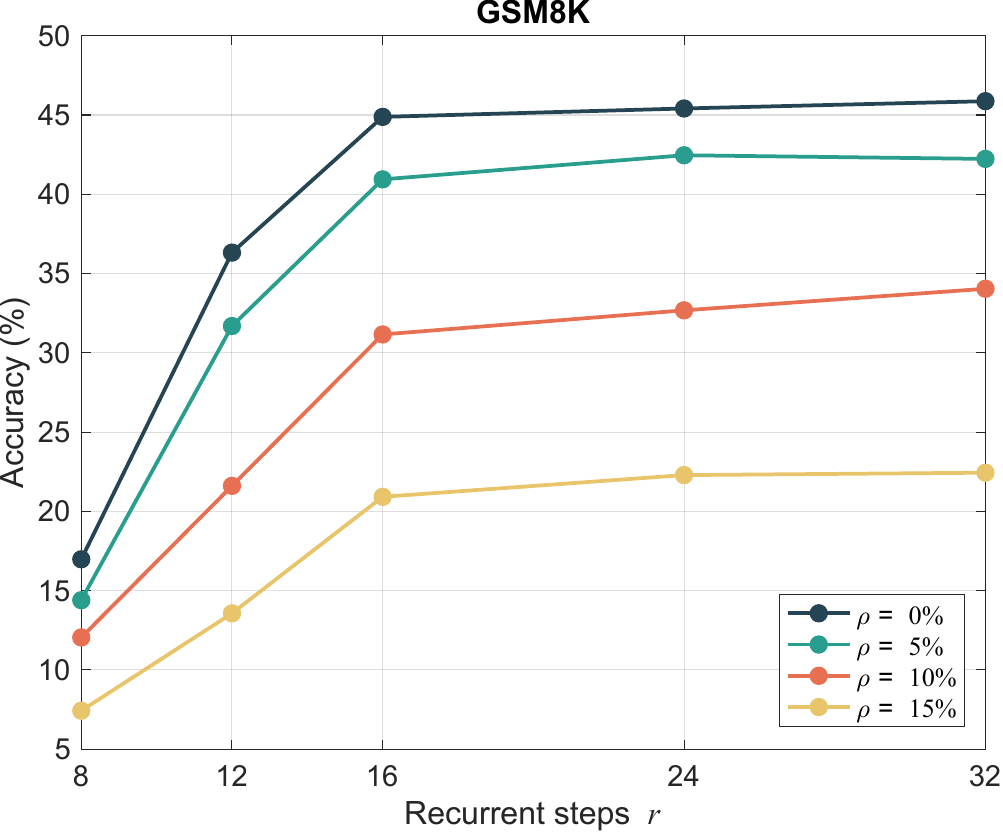}\label{fig: gsm8k_vs_recurrent}}
  \hfil
  \subfloat[MBPP accuracy vs. recurrent steps.]{%
    \includegraphics[width=0.32\textwidth]{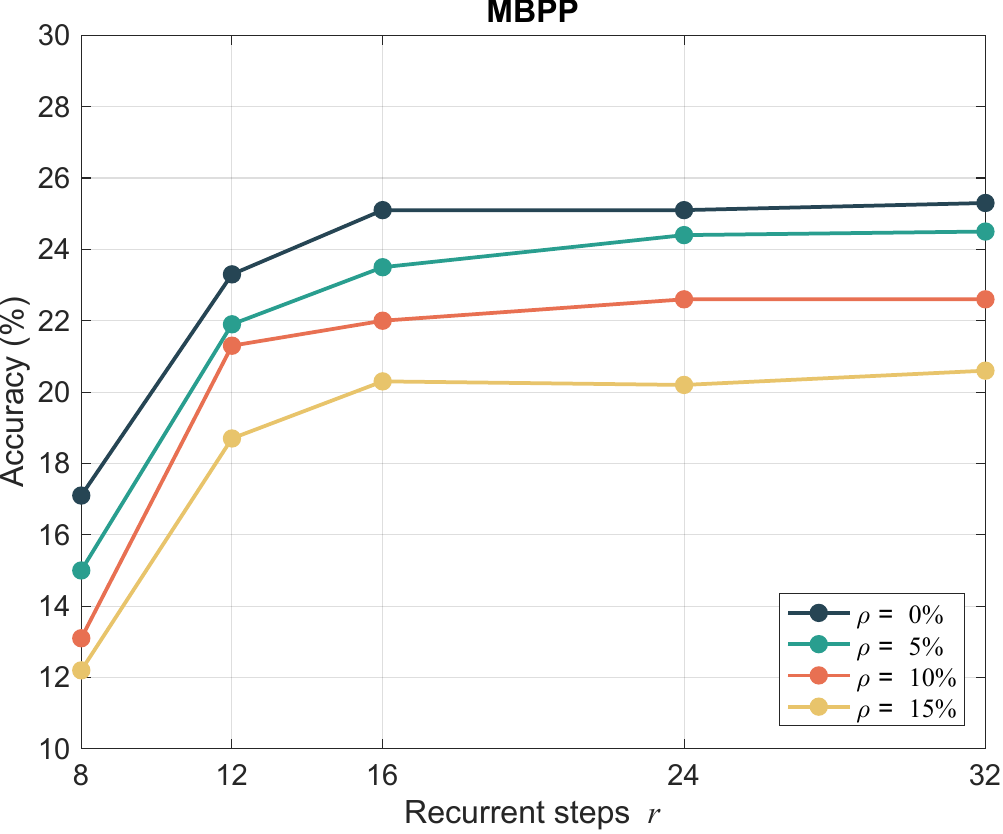}\label{fig: mbpp_vs_recurrent}}
  \hfil
  \subfloat[HellaSwag accuracy vs. recurrent steps.]{%
    \includegraphics[width=0.32\textwidth]{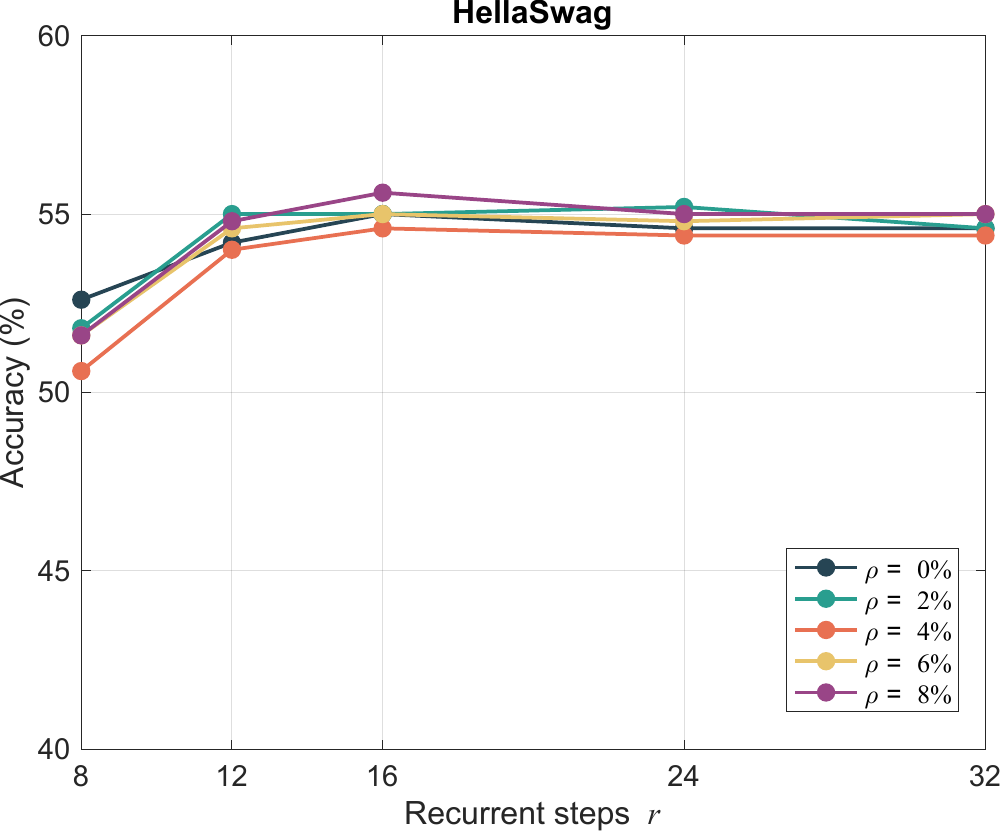}\label{fig: hella_vs_recurrent}}
  \caption{Accuracy vs. recurrent steps under different pruning rates.}
  \label{fig: insight_vs_recurrent}
\end{figure*}

Fig. \ref{fig: insight_vs_pruning} and Fig. \ref{fig: insight_vs_recurrent} jointly compare the accuracy trends of GSM8K, MBPP, and HellaSwag under different pruning rates and recurrent steps. For the generative tasks GSM8K and MBPP, two consistent patterns can be observed. First, for fixed recurrent steps, accuracy decreases as the pruning rate increases. Second, for a fixed pruning rate, accuracy improves as the recurrent steps increase, but the gain gradually diminishes and eventually saturates, which is more clearly illustrated in Fig. \ref{fig: insight_vs_recurrent}. However, the sensitivity to these factors varies significantly across datasets. In particular, MBPP is more resilient to the pruning rate. At $r=32$, increasing the pruning rate from $\rho=0$ to $\rho=15\%$ leads to only a 5\% accuracy drop, while GSM8K experiences an accuracy decline of 20\%. MBPP is also less sensitive to the recurrent steps, showing only a 2\% accuracy difference between $r=12$ and $r=32$, whereas GSM8K shows an accuracy difference of 10\%. This indicates that under resource-constrained settings, MBPP can be efficiently processed with fewer recurrent steps, while GSM8K requires deeper recurrent reasoning to maintain satisfactory performance. By contrast, HellaSwag, as a multiple-choice classification task, is much less sensitive to both the pruning rate and the recurrent steps. A moderate pruning rate yields negligible performance degradation, and the accuracy remains relatively stable as the recurrent steps increase. Overall, these results demonstrate that different tasks exhibit distinct response patterns to pruning rate and recurrent steps, indicating that resource allocation should be adaptively adjusted according to task characteristics.

\begin{figure*}[!t]
\centering
\includegraphics[width=16cm,height=5.5cm]{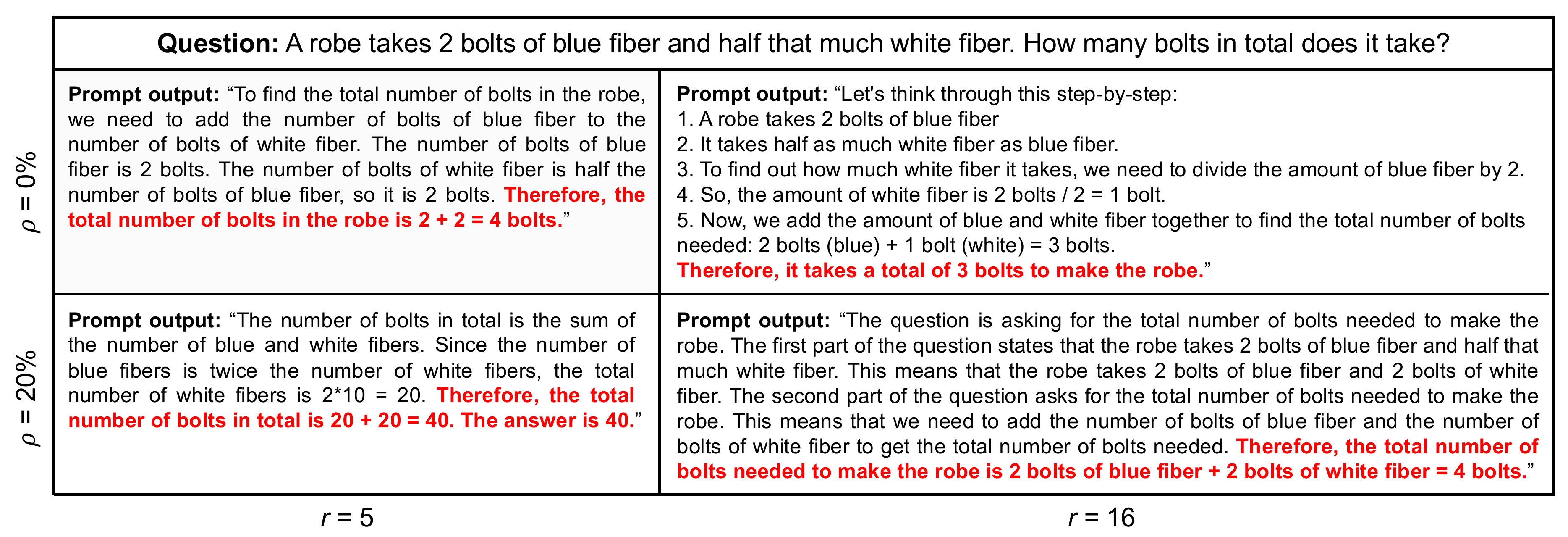}
\caption{Example GSM8K problem illustrating the effect of recurrent steps ($r$) and pruning rate ($\rho$) on reasoning outputs.}
\label{fig: gsm8k_example}
\end{figure*}

To provide an intuitive case-level perspective, Fig. \ref{fig: gsm8k_example} presents a representative GSM8K example evaluated under two pruning rates ($\rho=0$ and $\rho=20\%$) and two recurrent steps settings ($r=5$ and $r=16$), which clearly illustrates the impact of these parameters on reasoning performance. Only the configuration with $r=16$ and $\rho=0$ produces the correct answer with a coherent, step-by-step reasoning process. In all other settings, the reasoning becomes unreliable, with intermediate steps either omitted or incorrectly interpreted, ultimately leading to an incorrect output. This example aligns well with the trends in Fig.~\ref{fig: insight_vs_pruning} and Fig.~\ref{fig: insight_vs_recurrent}, showing that insufficient recurrent steps or aggressive pruning can compromise intermediate reasoning and thereby degrade final answer correctness. This further indicates that recurrent steps and pruning rate should be jointly selected to balance reasoning quality and resource efficiency.

\begin{figure*}[!t]
  \centering
  \subfloat[Device energy budget $E_{\max}^{\left({\mathrm{dev}}\right)}=60$.]{%
    \includegraphics[width=0.32\textwidth]{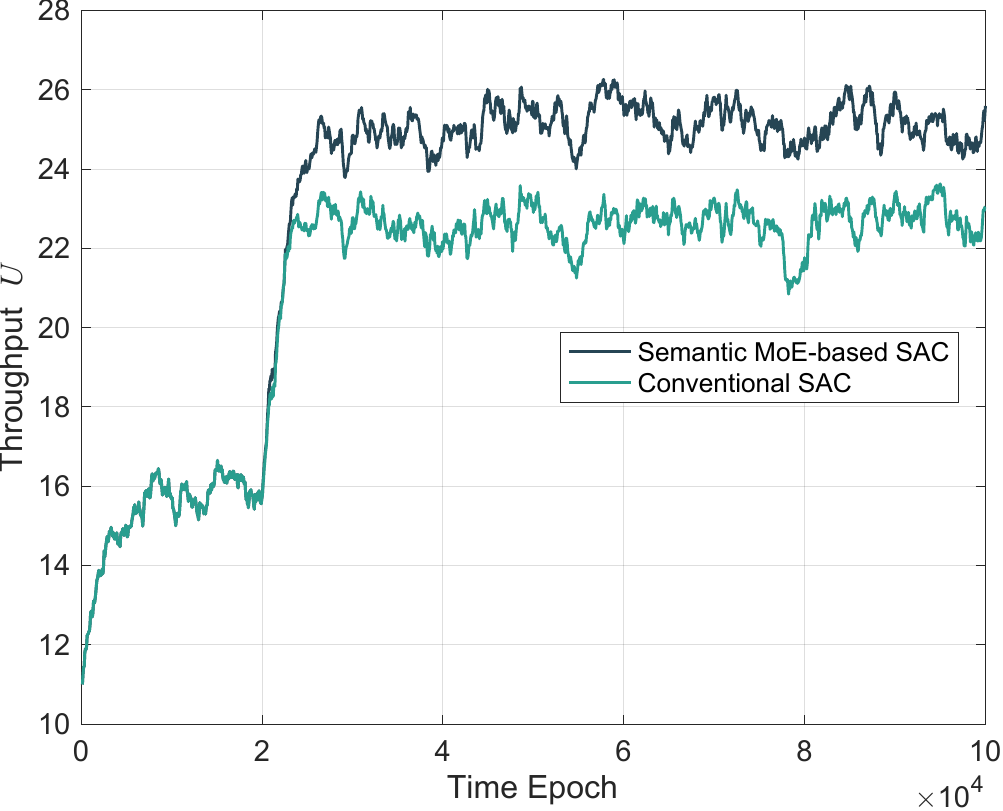}\label{fig: re4_s300k_d60}}
  \hfil
  \subfloat[Device energy budget $E_{\max}^{\left({\mathrm{dev}}\right)}=80$.]{%
    \includegraphics[width=0.32\textwidth]{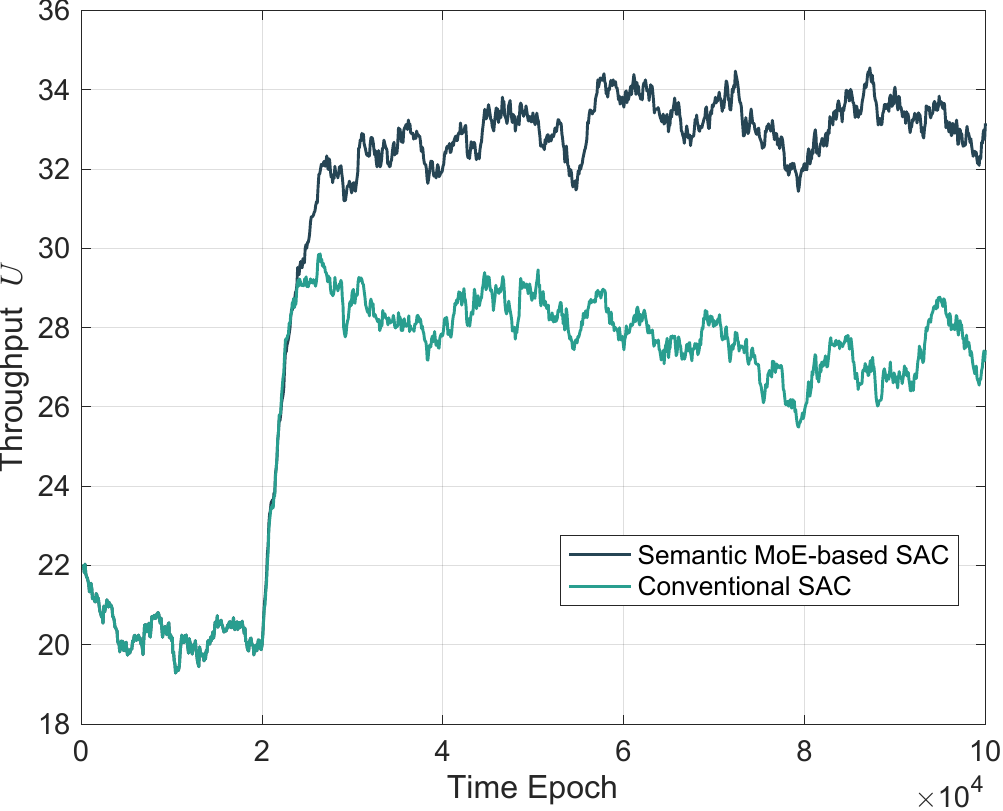}\label{fig: re4_s300k_d80}}
  \hfil
  \subfloat[Device energy budget $E_{\max}^{\left({\mathrm{dev}}\right)}=100$.]{%
    \includegraphics[width=0.32\textwidth]{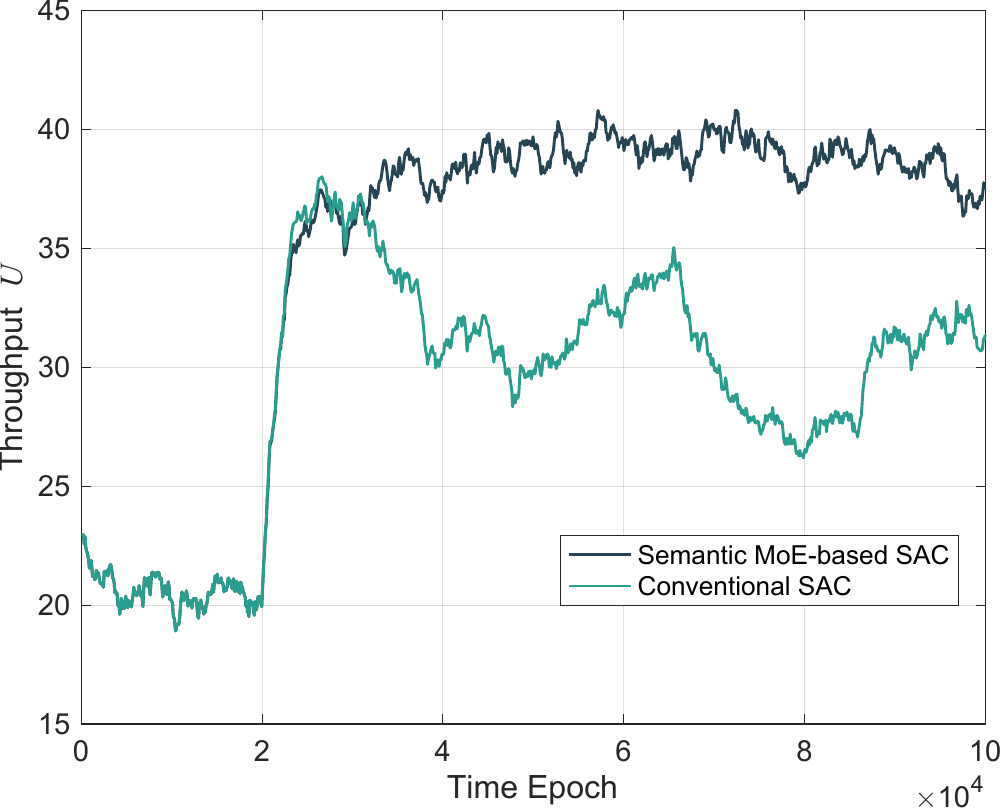}\label{fig: re4_s300k_d100}}
  \caption{Throughput over training epochs for MoE and baseline DRL models across different device energy budgets.}
  \label{fig: RL_device}
\end{figure*}

\begin{figure*}[!t]
  \centering
  \subfloat[Recurrence budget $G_{\max}=2 \times 10^5$.]{%
    \includegraphics[width=0.32\textwidth]{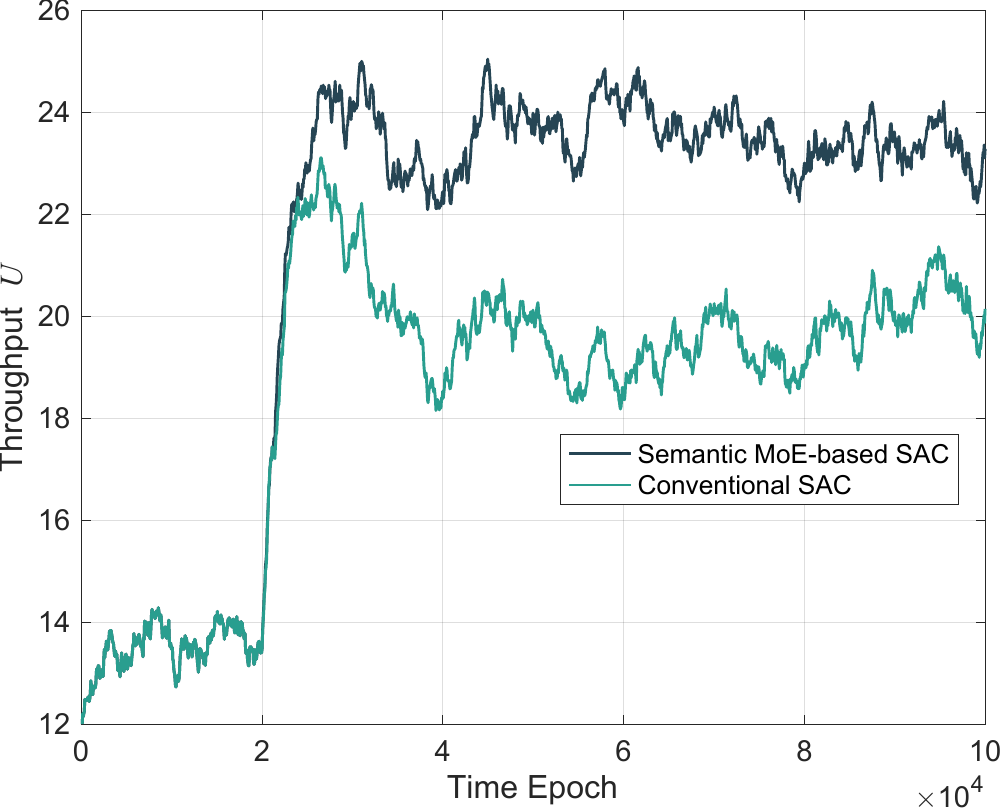}\label{fig: re4_s200k_d60}}
  \hfil
  \subfloat[Recurrence budget $G_{\max}=3 \times 10^5$.]{%
    \includegraphics[width=0.32\textwidth]{images/RL_re4_s300k_d60.pdf}\label{fig: re4_s300k_d60}}
  \hfil
  \subfloat[Recurrence budget $G_{\max}=4 \times 10^5$.]{%
    \includegraphics[width=0.32\textwidth]{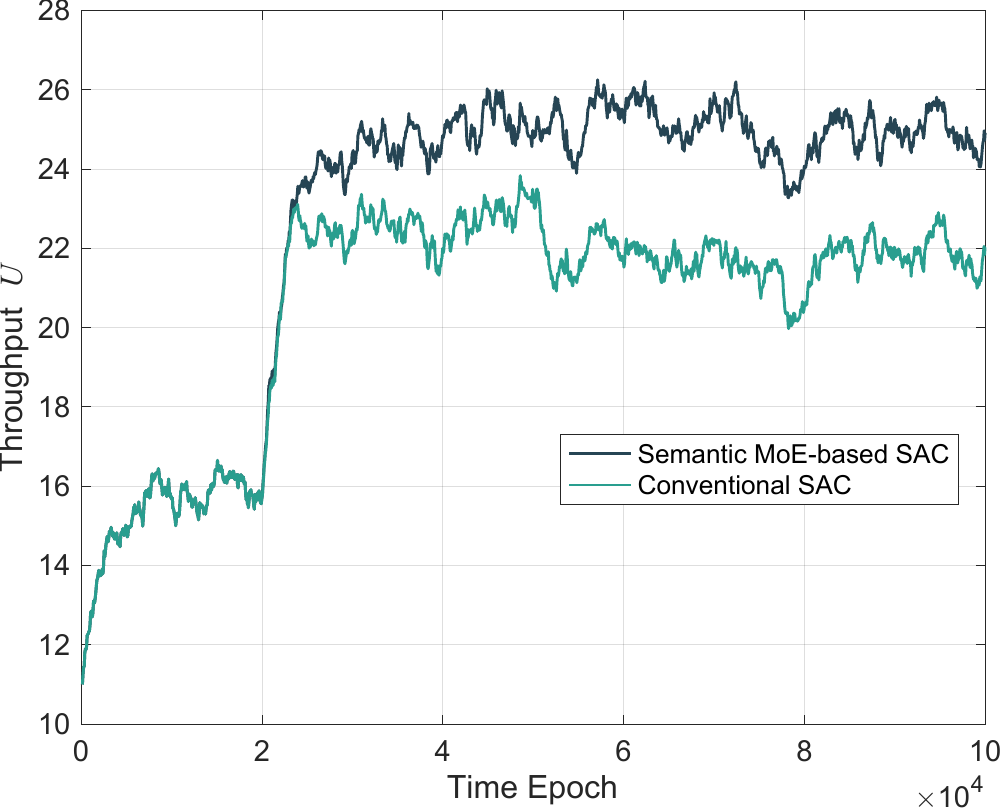}\label{fig: re4_s400k_d60}}
  \caption{Throughput over training epochs for MoE and baseline DRL models across different server recurrence budgets.}
  \label{fig: RL_server}
\end{figure*}

\begin{figure*}[!t]
  \centering
  \subfloat[Edge recurrent steps $r_{\text{e}}=0$.]{%
    \includegraphics[width=0.32\textwidth]{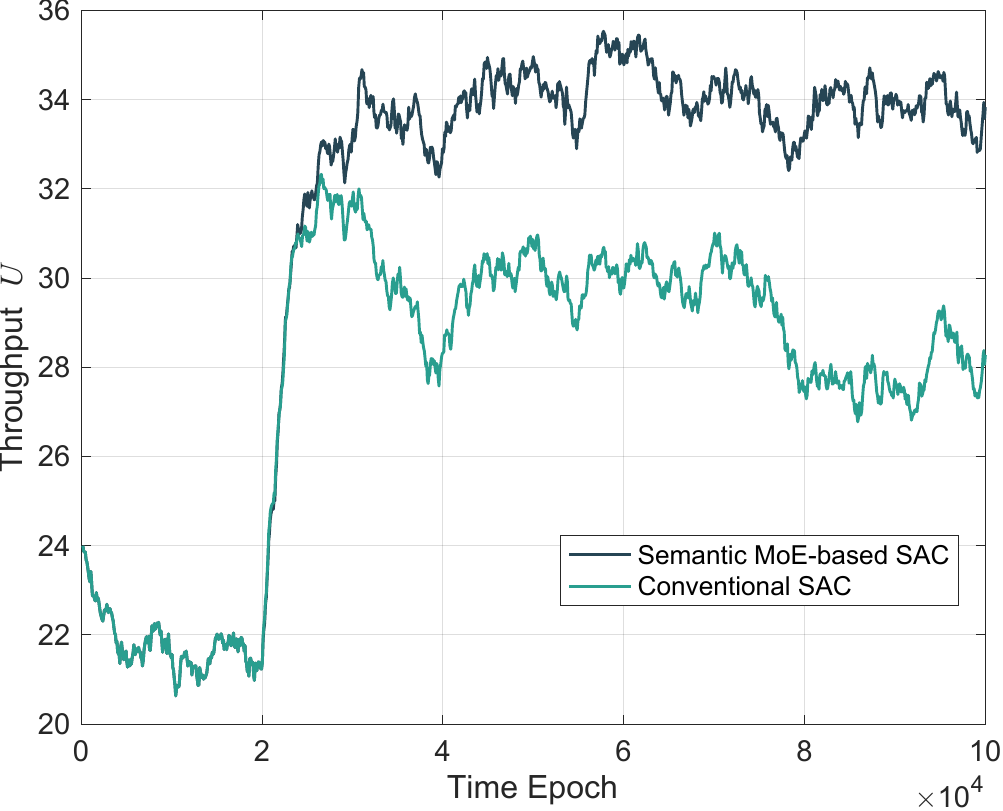}\label{fig: re0_s400k_d60}}
  \hfil
  \subfloat[Edge recurrent steps $r_{\text{e}}=2$.]{%
    \includegraphics[width=0.32\textwidth]{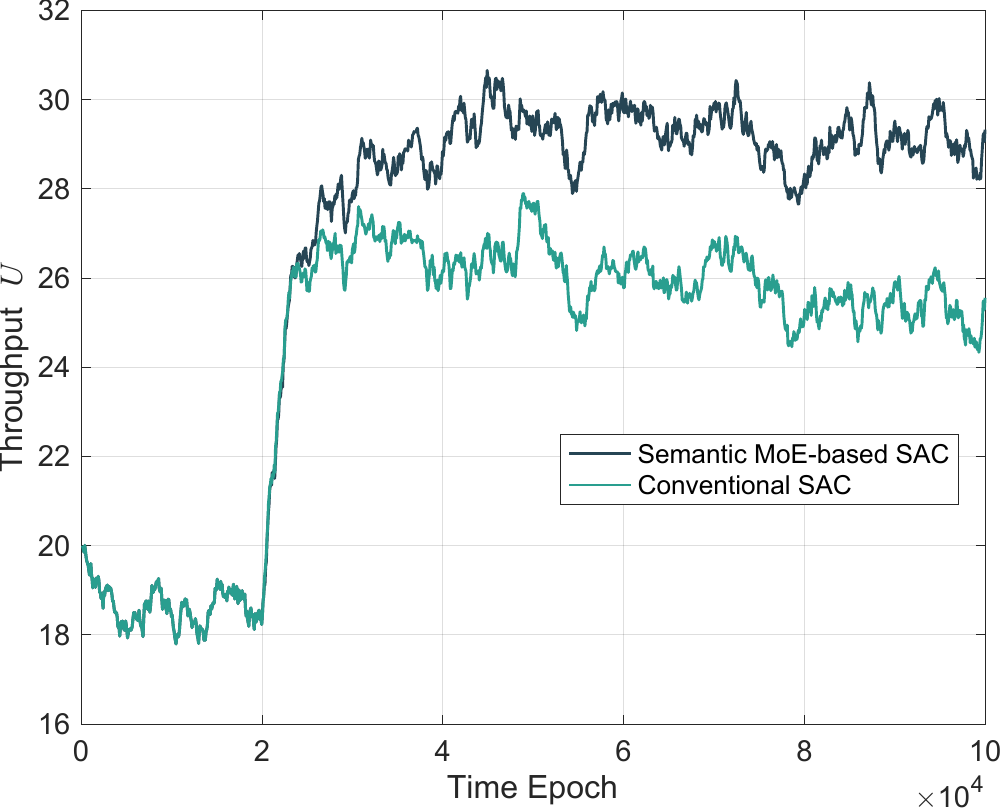}\label{fig: re2_s400k_d60}}
  \hfil
  \subfloat[Edge recurrent steps $r_{\text{e}}=4$.]{%
    \includegraphics[width=0.32\textwidth]{images/RL_re4_s400k_d60.pdf}\label{fig: re4_s400k_d60}}
  \caption{Throughput over training epochs for MoE and baseline DRL models across different initial edge iterations.}
  \label{fig: RL_re_distributed}
\end{figure*}

Fig. \ref{fig: RL_device} and Fig. \ref{fig: RL_server} compare the throughput of the proposed semantic MoE-based SAC and the conventional SAC under different resource constraints.
Specifically, Fig. \ref{fig: RL_device} evaluates performance with device energy budgets $E_{\max}^{\left({\mathrm{dev}}\right)}$ of 60, 80, and 100, while the recurrence budget $G_{\max}$ is fixed at $3 \times 10^5$. Conversely, Fig. \ref{fig: RL_server} considers $G_{\max}$ of $2 \times 10^5$, $3 \times 10^5$, and $4 \times 10^5$, with $E_{\max}^{\left({\mathrm{dev}}\right)}$ held constant at 60. For all these experiments, the number of edge recurrent steps is set to $r_{\text{e}}=4$.
As shown in both figures, the proposed semantic MoE-based SAC achieves stable throughput gains over the conventional SAC baseline across all settings. For instance, when $E_{\max}^{\left({\mathrm{dev}}\right)}=80$ and $G_{\max}=3 \times 10^5$, the proposed method achieves an average throughput of $U=33$ compared with $U=28$ for the baseline, yielding an improvement of 18\%.
As observed in Fig. \ref{fig: RL_device}, increasing the device energy budget results in higher throughput for the proposed method, and the performance gain over the baseline also increases. Specifically, as $E_{\max}^{\left({\mathrm{dev}}\right)}$ increases from 60 to 100, the average throughput of the proposed method rises from $U=25$ to $U=37$, while the corresponding gain increases from 9\% to 19\%.
A similar trend is observed in Fig. \ref{fig: RL_server}. In particular, as $G_{\max}$ increases from $2 \times 10^5$ to $3 \times 10^5$, the throughput rises from $U=23$ to $U=25$. However, further increasing the recurrence budget yields only marginal improvements, indicating that both methods are approaching saturation.
Although these budgets are sufficient for both models in the saturation region, our proposed method consistently outperforms the baseline due to the semantic MoE-based architecture.
Fig. \ref{fig: RL_re_distributed} further compares the two schemes across different edge recurrent steps $r_{\text{e}}$.
In this scenario, $E_{\max}^{\left({\mathrm{dev}}\right)}$ and $G_{\max}$ are fixed at 60 and $4 \times 10^5$, respectively.
Generally, a smaller $r_{\text{e}}$ reduces device energy consumption but increases server recurrence usage for robust reasoning. As shown in the figure, our proposed method consistently performs better than the baseline regardless of $r_{\text{e}}$. We also observe that as $r_{\text{e}}$ increases, the average throughput decreases from $U=34$ to $U=25$. Since the device energy budget is the primary constraint in this setting, a higher $r_{\text{e}}$ consumes more energy at the device and thus results in lower average throughput as expected.

\begin{figure*}[!t]
  \centering
  \subfloat[GSM8K with poor channel]{\includegraphics[width=0.3\textwidth]{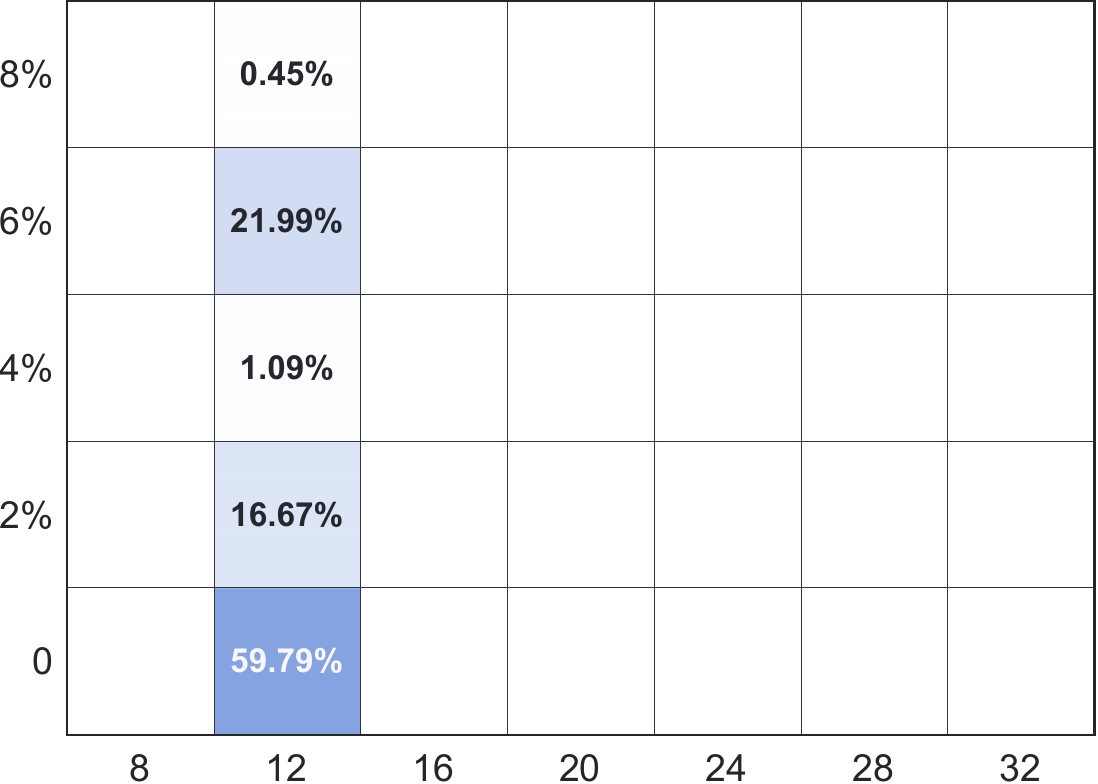}}\hfill
  \subfloat[GSM8K with fair channel]{\includegraphics[width=0.3\textwidth]{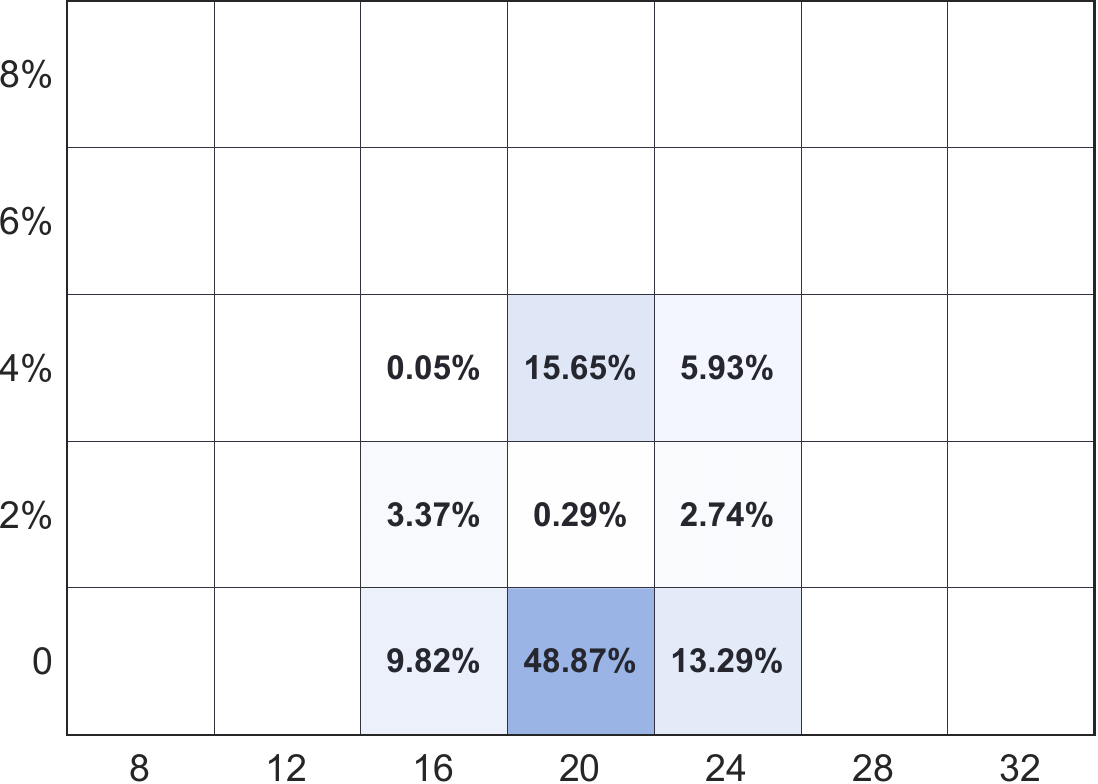}}\hfill
  \subfloat[GSM8K with good channel]{\includegraphics[width=0.3\textwidth]{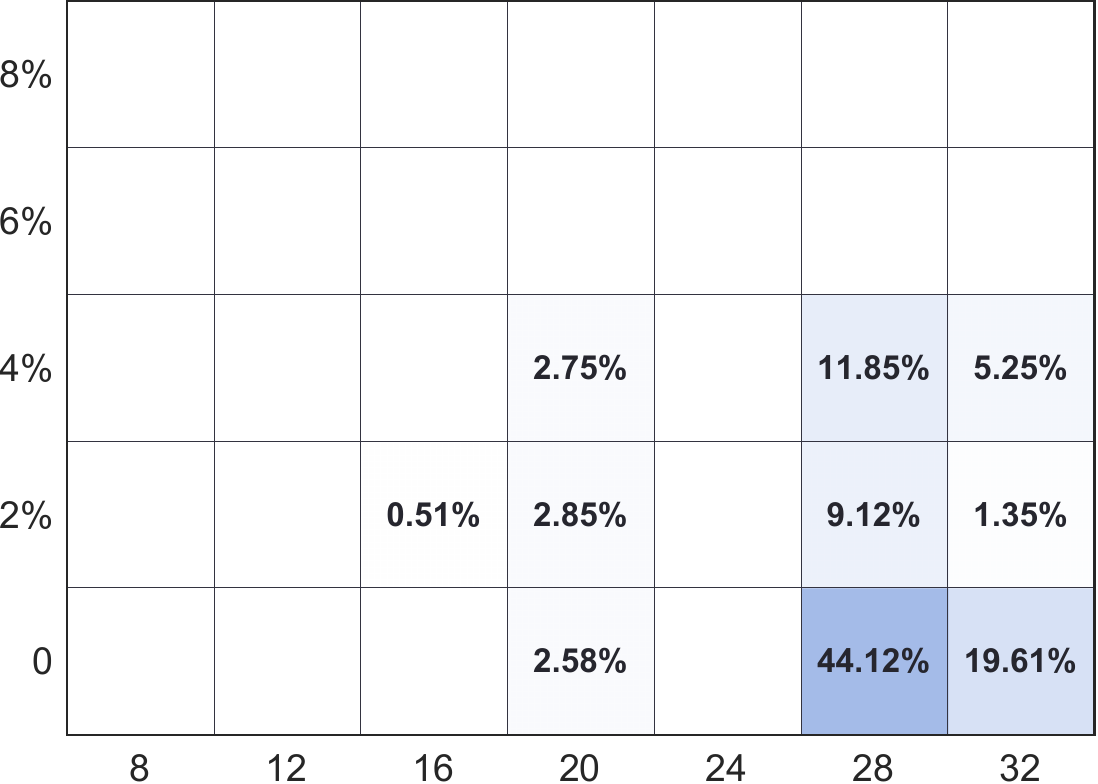}}\\[2pt]
  \subfloat[MBPP with poor channel]{\includegraphics[width=0.3\textwidth]{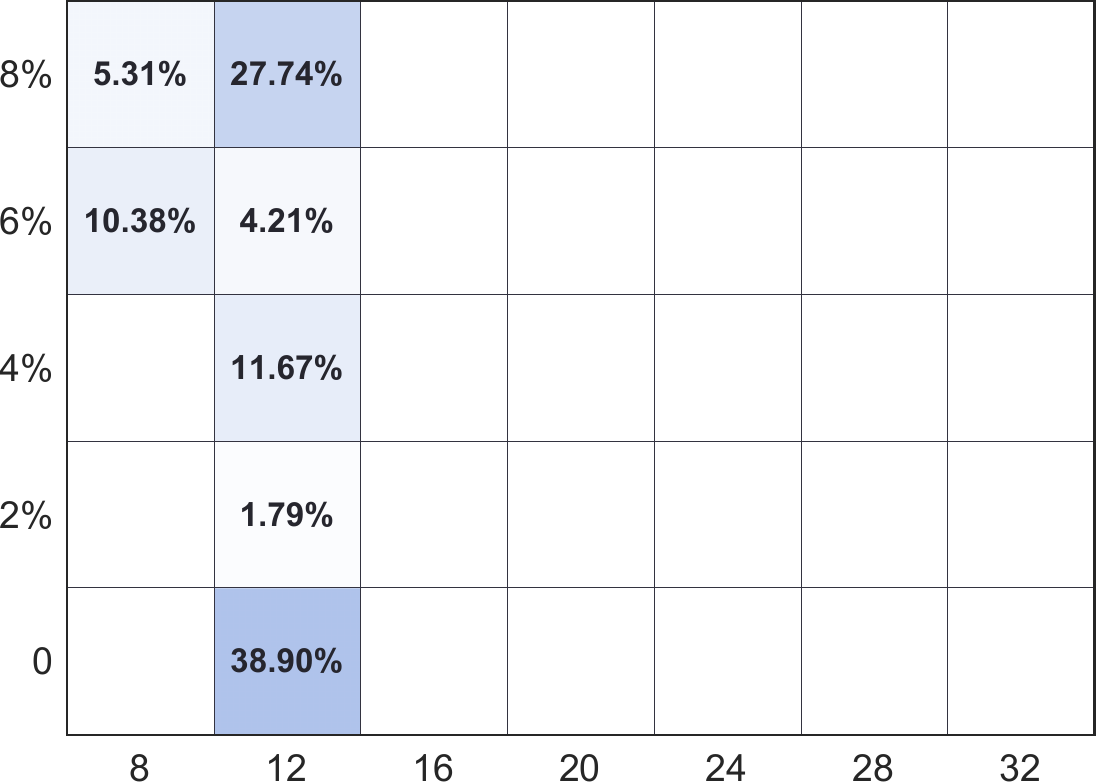}}\hfill
  \subfloat[MBPP with fair channel]{\includegraphics[width=0.3\textwidth]{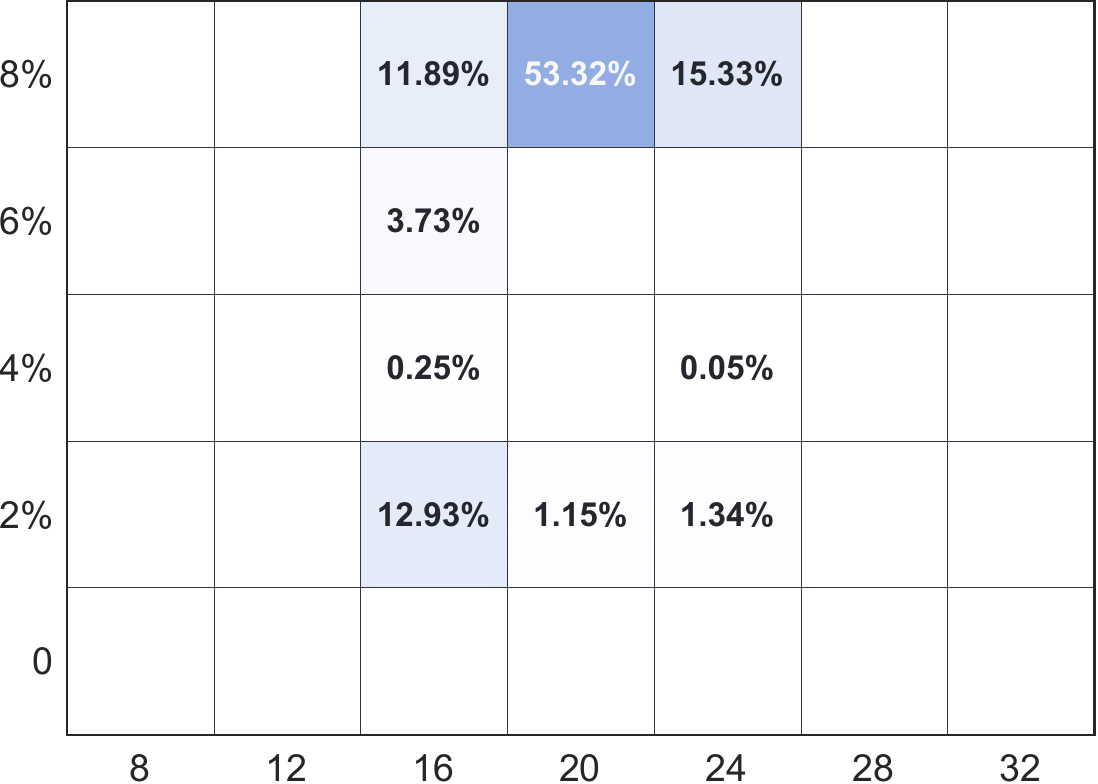}}\hfill
  \subfloat[MBPP with good channel]{\includegraphics[width=0.3\textwidth]{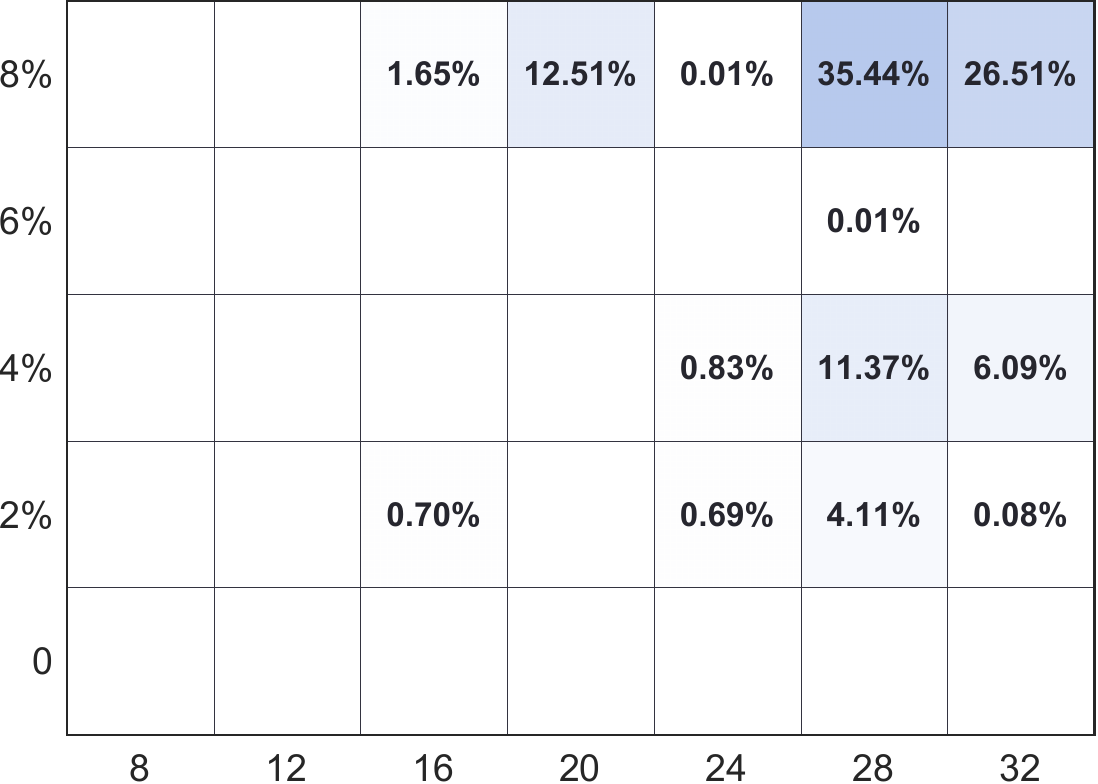}}\\[2pt]
  \subfloat[HellaSwag with poor channel]{\includegraphics[width=0.3\textwidth]{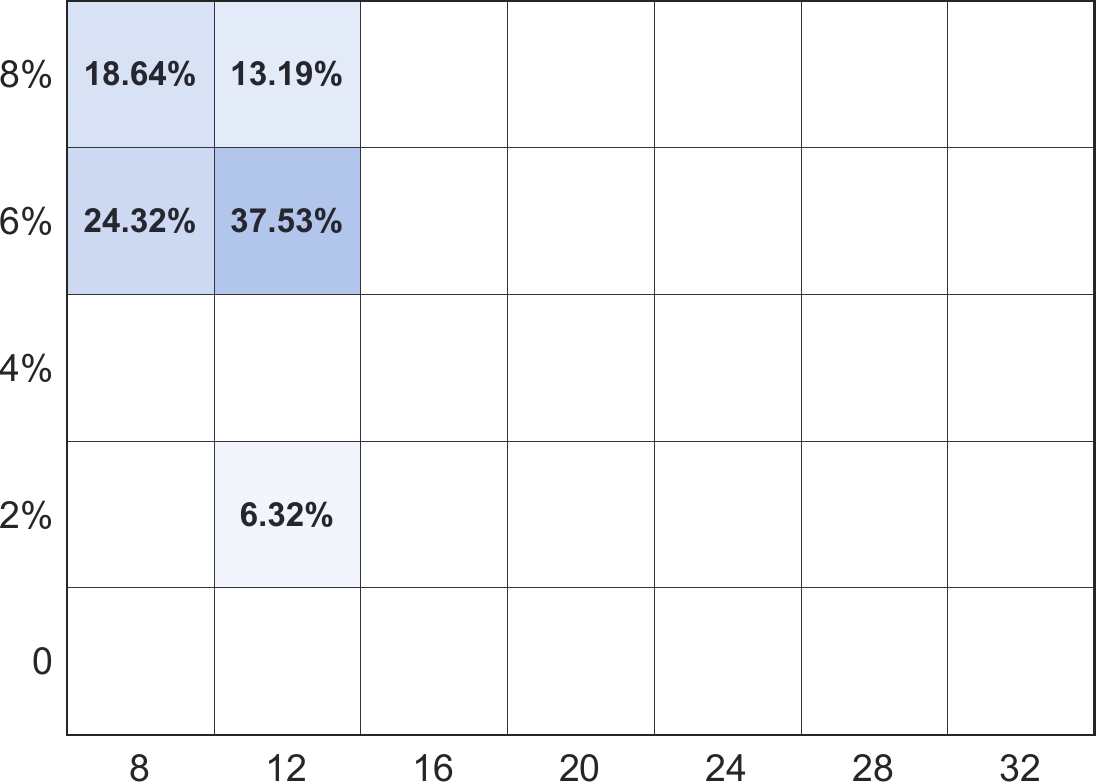}}\hfill
  \subfloat[HellaSwag with fair channel]{\includegraphics[width=0.3\textwidth]{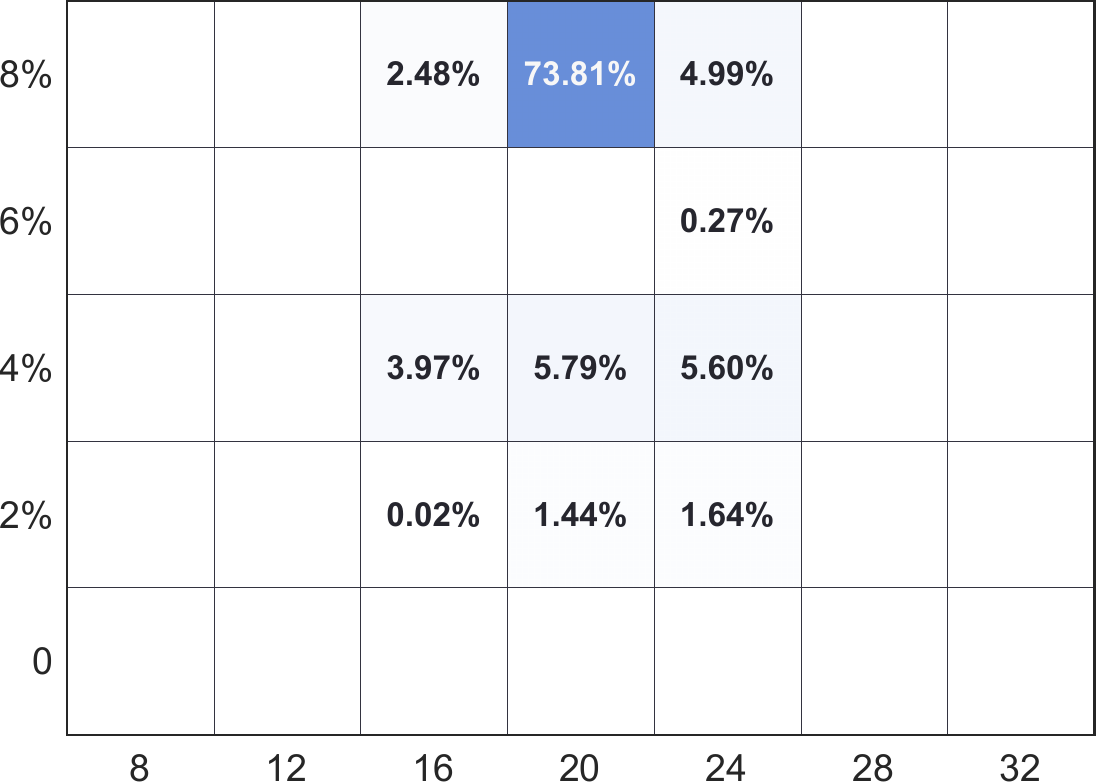}}\hfill
  \subfloat[HellaSwag with good channel]{\includegraphics[width=0.3\textwidth]{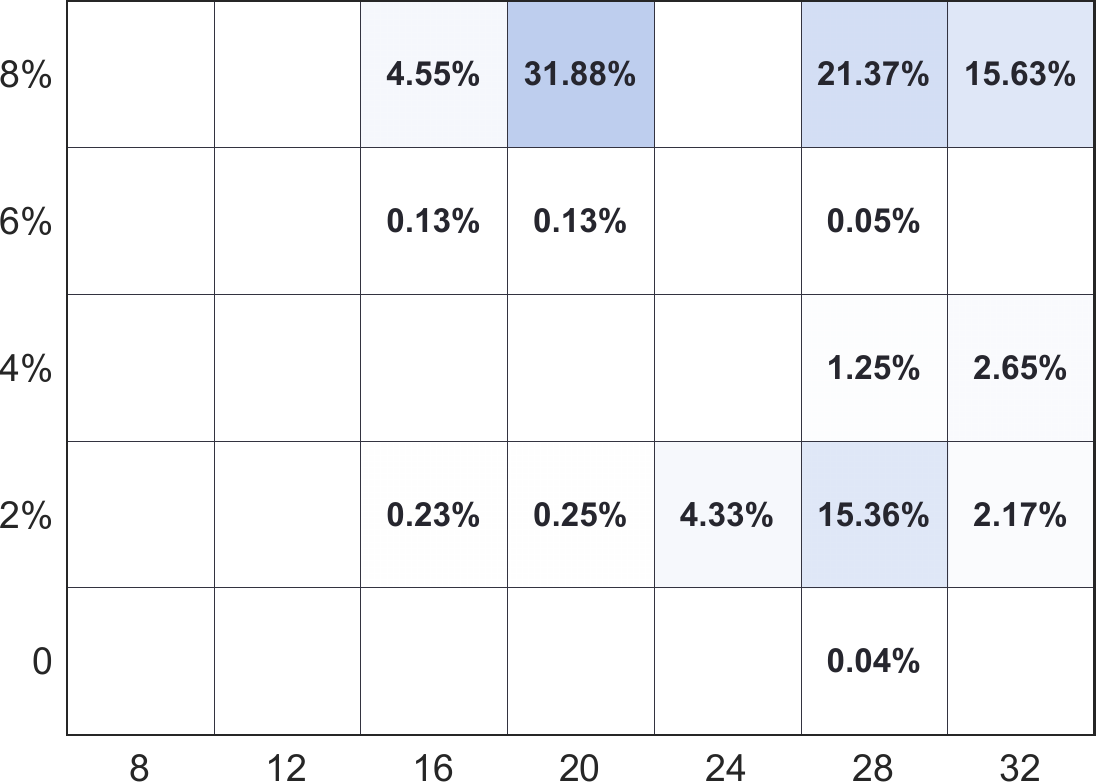}}
  \caption{Expert decision distributions across distinct task and channel conditions.}
  \label{fig: expert_heatmap}
\end{figure*}

Fig. \ref{fig: expert_heatmap} illustrates the expert decision distributions across distinct task and channel conditions.
Each sub-figure corresponds to a specific task–channel scenario, and the heatmap shows the probability of selecting different combinations of the recurrent steps and the pruning rate. The results indicate that each expert learns a distinct strategy under different task and channel conditions, which can be explained as follows. As indicated in Fig.~\ref{fig: insight_vs_pruning} and Fig.~\ref{fig: insight_vs_recurrent}, different tasks exhibit distinct sensitivities to the pruning rate and the recurrent steps. Such task-dependent sensitivities are directly reflected in the expert strategies in Fig.~\ref{fig: expert_heatmap}. Specifically, since GSM8K is more sensitive to the pruning rate, the corresponding experts tend to choose a smaller pruning rate, whereas MBPP and HellaSwag tend to choose a larger pruning rate to reduce communication energy consumption. Meanwhile, GSM8K benefits more from larger recurrent steps, followed by MBPP and then HellaSwag. This is captured by the higher probabilities of selecting larger recurrent steps for GSM8K under the same channel conditions. In addition, for a fixed task, better channel conditions lead experts to prefer larger recurrent steps, allowing the system to exploit higher channel quality for improved throughput.

Table~\ref{table: semantic_bar} compares the average throughput under different routing strategies. 
We compare three settings, including an oracle baseline with ground-truth task types, the proposed semantic router scheme that predicts task types via classification, and a random assignment baseline without task information. The results show that the proposed semantic router achieves throughput close to the oracle baseline, indicating that it can accurately predict task types and effectively approximate oracle-level performance. In contrast, random assignment results in a 5\% throughput reduction, demonstrating the importance of exploiting task semantics for expert selection. Moreover, this gain is expected to be more significant as task diversity increases.

\begin{table}[h]
    \centering
    \caption{Comparison of average system throughput under different routing strategies.}
    \label{table: semantic_bar}
    \begin{tabular}{l|c}
        \toprule
        \textbf{Routing Strategy} & \textbf{Average Throughput $U$} \\
        \midrule
        Oracle   & 30.1 \\
        Semantic & 30.0 \\
        Random   & 28.4 \\
        \bottomrule
    \end{tabular}
\end{table}

\section{Conclusion}\label{sec: conclusion}
We proposed MORES, an LLM reasoning framework that enables distributed inference-time scaling in wireless edge networks.
By leveraging recursive latent reasoning, MORES supports natural computation partitioning between edge devices and edge servers, allowing devices to access on-demand reasoning depth under heterogeneous resource constraints.
To ensure robust long-term performance under dynamic wireless conditions, we formulated a joint computation–communication scheduling problem and solved it through a semantic MoE-based DRL scheme. 
Experimental results demonstrate that the proposed DRL solution effectively handles the heterogeneity of the system, achieving consistent performance gains over representative baselines.
In future work, we will extend MORES to incorporate multimodal reasoning capabilities and network-economic mechanisms in real-world wireless deployments.

\bibliographystyle{IEEEtran}
\bibliography{Ref}

\end{document}